\documentclass[11pt]{article}  

\usepackage{amsmath,amsthm,latexsym,amssymb,amsfonts,epsfig}

\newtheorem{Theorem}{Theorem}[section]

\newtheorem{Lemma}{Lemma}[section]

\newcommand{\be}{\begin{equation}}
\newcommand{\ee}{\end{equation}}
\newcommand{\ba}{\begin{eqnarray}}
\newcommand{\ea}{\end{eqnarray}}

\title{{\sf Non-perturbative, background independent canonical quantum gravity
in Fock representations}}
\author{
{\sf T. Thiemann}$^1$\thanks{{\sf 
thomas.thiemann@gravity.fau.de}}\\
\\
{\sf $^1$ Inst. for Quantum Gravity, FAU Erlangen -- N\"urnberg,}\\
{\sf Staudtstr. 7, 91058 Erlangen, Germany}\\
}
\date{{\small\sf \today}}

\makeatletter
\@addtoreset{equation}{section}
\makeatother

\begin{document} 

\maketitle

{\sf

\begin{abstract}
It is commonly believed that a quantum field theory of General Relativity requires 
a non-perturbative formulation. In addition, the background independence of classical General 
Relativity supplies a physical selection criterion for suitable Hilbert space representations 
of the corresponding quantum field theory.

In this contribution we show that there exist background independent representations 
of Fock type within the manifestly non-perturbative, canonical approach to quantum gravity.
Mandatory for their existence is the presence of suitable matter fields next to the geometry 
field. In particular, the excitations of the corresponding Fock vacuum necessarily entangles matter and 
geometry.

In this article we use the constraint quantisation method. We compare the resulting Fock incarnation 
of background independent, non-perturbative canonical quantum gravity with the well known Loop quantum 
gravity incarnation. One of the most important differences is that the Fock quantum gravity (FQG) Hilbert space, 
in contrast to the Loop quantum gravity (LQG) Hilbert space, is separable. This has many 
advantages when attempting to implement the Hamiltonian constraint, also known as Wheeler-DeWitt constraint, 
as a densely defined quadratic form.    
\end{abstract}

\section{Introduction}
\label{s1}

It is widely believed that perturbative (PT) quantum gravity (QG) is a non-renormalisable
quantum field theory (QFT) \cite{1}. Accordingly many approaches to quantum gravity attempt 
a non-perturbative (NP) quantisation of General Relativity (GR) (gravity and matter)
\cite{2,3,4,5,6}. Among those, the canonical quantum gravity (CQG) approach is historically the oldest
\cite{7,8,9,10,11} but despite tremendous effort it was very hard to put the theory on 
sound mathematical footing. This changed with the introduction of connection variables 
for the gravitational field \cite{12,13} and the mathematical theory invented for them
(see e.g. \cite{6}) and references therein). Today we have a precise mathematical 
formulation of the canonical quantisation approach to quantum gravity and Loop 
quantum gravity (LQG) is a concrete incarnation of that programme. 

While LQG has not 
yet been completed, important steps have been taken. Two of the most important steps 
were the choice of 1. a suitable $^\ast-$algebra $\mathfrak{A}$ of operators and 2. a suitable Hilbert space 
representation of the corresponding canonical synchronous (anti) commutation 
and adjointness relations among elements of $\mathfrak{A}$. For a cyclic representation, one can encode the 
latter choice via a state $\omega$ on $\mathfrak{A}$ using the Gel'fand-Naimark-Segal (GNS) construction
\cite{14}. Thus CQG uses elements of the algebraic approach (AQFT) to QFT. It does not use all the structure 
of AQFT because the most important axiom of AQFT is locality or causality which requires that two operator 
valued distributions smeared with test functions of causally disconnected support (anti-)commute. 
This very statement needs as an input a classical spacetime metric with respect to which one can determine the 
lightcone structure of the spacetime manifold. As in QG the metric itself is quantised, the locality 
requirement cannot be adopted as stated in AQFT. In QG {\it all} spacetime metrics must be incorporated not 
only a single one, these are fluctuating operators rather than a fixed classical background structure
and lightcones become fuzzy. Locality has to be reformulated,
perhaps as a corresponding statement in terms of matter operators smeared over spacetime 
manifold regions in partial expectation values with respect 
to gravitational state vectors that are semiclassically peaked on a classical spacetime metric.

This crucial difference between CQG and (A)QFT on fixed (curved) spacetime (CST) metrics reflects 
an important feature of classical GR called background independence (BI). Loosely speaking this 
means that Einstein's theory a priori does not designate a special role to a distinguished 
spacetime metric, there are infinitely many metrics, even on shell (i.e. solutions to Einstein's equations)
and in QG we can e.g. superpose, say two semiclassical states peaked on two very different classical spacetime metrics 
which are not perturbations of each other. Because of this feature, the classical GR action is invariant under 
the gauge symmetry of spacetime diffeomorphisms and the corresponding Euler-Lagrange equations (Einstein's equations)
are covariant. One often hears the statement that non-perturbative treatments of QG imply background 
indepence. However, that is not quite true. What is correct is that a perturbative treatment based on 
splitting the full metric into a background and a perturbation (called the graviton), perturbing 
the gravitational action around that background and then performing 
(A)QFT for the resulting graviton theory on that background (often chosen as Minkowski metric) 
necessarily is backround dependent (BD). As mentioned above, this procedure leads to a perturbatively 
non-renormalisable and therefore non-predictive theory. However, it is perfectly possible to use background dependent algebras 
and states while treating the gravity action non-perturbatively. This elementary point, 
namely that the logical negation of PT $\Rightarrow$ BD is BI $\Rightarrow $ NP but does not 
imply NP $\Rightarrow$ BI has been emphasised recently in \cite{16,17} where new possibilities 
in the choice of $\mathfrak{A},\omega$ for NP CQG were explored, both BI and BD ones. 
In fact, these two papers form the foundation of the present work.   
   
Yet, the principle of background independence can play another important role, in particular
in relation to the afore mentioned predictivity of the theory. This has to do 
with the fact that in QFT the set of states $\omega$ on given $\mathfrak{A}$ is typically
uncountably infinite even when restricting to pure (the representation is irreducible)
and regular (one can pass from the Weyl to the Heisenberg algebra) ones. The corresponding GNS representations of $\mathfrak{A}$ are 
typically unitarily inequivalent leading to different spectra for self-adjoint (observable)
elements. In QFT one therefore must invoke additional physical input in order to select 
a distinguished class of states and hope that they depend on only a finite number 
of parameters. Typically, this encodes symmetries and/or dynamical information
such as Poincar\'e invariance or that the Hamiltonian is implementable as a self-adjoint operator.
In QG one may thus use BI to select states and this is in fact what has been done on LQG 
where one could prove that for the choice of $\mathfrak{A}$ made there the (spatially) diffeomorphism
invariant states are unique. Spatial diffeomorphisms can therefore be 
represented by unitary operators. 

Part of motivation for \cite{16,17} and the present work are particular features of that unique 
LQG state, namely that the corresponding GNS Hilbert space is not separable and that it is 
not strongly continuous with respect to one-parameter unitary subgroups of certain 
Weyl subgroups and the spatial diffeomorphism
group. This and related consequences have made it quite difficult to implement the so-called 
hypersurface deformation algebroid (HDA). 
As has been shown in \cite{18}, this is a universal structure of canonical formulations 
of classical theories descending from a covariant spacetime Lagrangian (at most second order in derivatives
to avoid the Ostrogradski instability \cite{19}) . 
Here the term ``algebroid'' accounts for the fact that this structure is not a Lie algebra. 
For the canonical 
formulation of GR the HDA is represented in terms of Poisson bracket relations among the generators of 
spatial and temporal spacetime diffeomorphisms and their (anomaly free) implementation in the quantum theory
is therefore an important necessary ingredient in any CQG incarnation. The fact that in LQG the generator 
of spatial diffeomorphisms cannot be defined due to the afore mentioned lack of continuity (Stone's theorem
cannot be used) makes it clear that a representation of the HDA will encounter difficulties. 
While several solutions to this problem have been proposed and partly implemented, the issue has not been
completely resolved yet, see \cite{20} for a recent review. Perhaps the most promising and 
most practical solution is to abandon the quantisation of the constraints and rather solve them
classically using gauge fixing (or equivalently relational Dirac observables). This so-called 
reduced phase space approach has been implemented in LQG using various types of matter, see e.g.
\cite{21} and references therein. Yet, the non-separability of the Hilbert space remains and the Weyl discontinuity implies 
quantisation ambiguities which one may hope to fix using ideas from renormalisation, see e.g. \cite{22}
and references therein.

Another consequence of non-separability 
is that it is quite difficult to define semi-classical states. This is because in LQG the GNS vacuum 
can only be excited by operators supported on 1-dimensional submanifolds (more specifically, finite graphs)
and as we can superimpose at most a countable number of them to obtain a normalisable state, even such 
excited states are degenerate almost everywhere in the sense that they are zero eigenstates 
of the LQG quantum volume operator at an uncountably infinite number of points of non-vanishing 
Lebesgue measure. As the {\it inverse} 
of the spatial metric enters the HDA this feature presents another source of problems towards 
its quantum implementation \cite{23}. While it is possible 
to define normalisable LQG states in LQG for which the volume eigenvalue is positive for any open region
(see appendix of \cite{24}) the semiclassical properties of such states have not been explored yet.\\
\\
In the present paper and in \cite{16,17} we therefore ask whether a different CQG incarnation is possible which avoids 
non-separability and the consequences that it has in the LQG incarnation. In \cite{16,17} we discussed 
both BD and BI implementations of this idea. Moreover, there we considered both the constraint quantisation 
and reduced phase space quantisation route. In the present paper we develop this further, focussing 
on non-perturbative and background independent implementations of quantum constraints. More specifically
we ask for choices of $\mathfrak{A},\omega$ which lead to NP and BI Fock representations. Surprisingly,
this is indeed possible, if one admits a sufficient amount of (standard model) matter and imposes 
certain inequalities on the classical phase space such as non-degeneracy which is what is silently done 
in most approaches to QG. Of course such classical inequalities are typically broken in the quantum theory
in states which are very far from being semi-classical but this is mathematically and logically allowed.
On the other hand, NP and BI Fock representations for vacuum GR (without matter) appear to be impossible
if they are meant to be practically useful.
We explore the properties of this new incarnation and compare with the LQG incarnation in particular 
with respect to the following aspects: 1. frame rotation and spatial diffeomorphism covariance, 2.
existence of constraint operators respectively quadratic forms, 3. quantum HDA implementation,
4. existence of geometrical operators or quadratic forms (length, area, volume in 3+1 dimensions), 
5. semiclassical, concretely coherent, states, 6. kernel of the quantum constraints as generalised
(not normalisable) zero eigenvectors and 7. Hilbert space structures on that kernel.\\
\\
The organisation of this work is as follows:\\
\\

In section \ref{s2} we review the canonical quantisation programme of constrained field theories 
and then summarise its application to GR in the LQG incarnation in order to compare later to 
analogous structures in the Fock incarnation.

In section \ref{s3} we derive the general structure of BI Fock algebras and states and then 
show how these can be found for GR including sufficient amount of (standard model) matter.
An important lesson is the neccessity to work with scalars of fractional density weight 
$\frac{1}{2}$.

In section \ref{s4} we show that in the corresponding Fock representations the Gauss and 
spatial diffeomorphism constraints can be implemented as self-adjoint operators while 
the Hamiltonian constraint or the corresponding 
master constraint \cite{25} (in a certain density weight) exists as a densely defined 
symmetric quadratic form. Crucial for this is of course the natural normal ordering
with respect to the the available Fock structure.

In section \ref{s5} we show that the sub-Lie-algebra of the HDA generated by 
Gauss and spatial diffeomorphism constraints closes without anomalies and that their 
commutator with the Hamiltonian or master constraint quadratic form closes in the expected way.
The commutator of Hamiltonian constraint quadratic forms is ill defined without 
regularisation as quadratic forms cannot be multiplied. However, at least for 
Euclidian signature, in similar but BD Fock representation a regularisation has been found 
such that the HDA is successfully implemented \cite{17}. Work is in progress in order 
to apply these ideas also in the present incarnation.

In section \ref{s6} we show that simple rigorous solutions to both the Gaus and spatial 
diffeomorphism constraints exist as generalised zero eigenvectors. Generalised zero 
eigenvectors have the advantage that the equations they obey only require the 
constraints to be quadratic forms, hence these ideas can also be applied to the 
Hamiltonian constraint although of course it is much more complicated to find exact
solutions than for the other constraints.

In section \ref{s7} we show that the Fock Hilbert space admits standard coherent states 
as semiclassical states which saturate the Heisenberg uncertainty bound for 
the Fock annihilation and creation operators. We also introduce domains based 
on coherent states in order to quantise geometrical quantities. We also show how 
one can in principle derive the LQG state as an infinitely squeezed Fock state.

In section \ref{s8} we construct geometrical quantities as quadratic forms. Here we make 
use of ideas of Weyl and deformation quantisation \cite{26,27}. These techniques are also 
necessary in order to define solutions to the constraints beyond the simple ones discussed 
in section \ref{s6}. An important property of these quadratic forms is that they are 
densely defined not on the span of Fock states (finite particle excitations of the Fock vacuum) but 
rather e.g. the span of finite excitations of a coherent state whose metric (or D-Bein in D+1 dimensions)
expectation value is non-degenerate. Here we use the well known fact that Fock states are 
pure \cite{28} so that every vector in the GNS Hilbert space is cyclic.    

In section \ref{s9} we summarise and conclude. At this point we mention a much shortened
version of the present paper \cite{28a} which focuses on structure rather than detailed 
proofs and thus could be consulted on a first reading.

\section{Canonical quantisation of constrained systems and the LQG incarnation}
\label{s2}

In the first subsection we sketch the canonical quantisation of constrained systems 
and in the second we summarise its application to GR in the LQG incarnation of CQG.              
The presentation will be short, all the details can be found in \cite{5}.

\subsection{Canonical quantisation of constrained systems}
\label{s2.1}
 
We restrict the presentation to bosonic fields, the discussion for fermionic fields is 
different \cite{16,17} but does not present additional obstacles. We work with finite 
or countably infinite discrete index sets which is always possible upon smearing 
functions with respect to a countable basis of test functions.\\  
\\
We consider a phase space with real valued canonically conjugate fields $(q^a,p_a)$ that is, the 
non-vanishing Poisson brackets are $\{p_a,q^b\}=\delta_a^b$. Here we assume for the moment that 
both $q,p$ take values in a vector space. This assumption will be relaxed later.
We also assume that the
system is totally constrained by real valued first class functions $C_I=C_I(q,p)$ which means 
that $\{C_I,C_J\}=\kappa_{IJ}\;^K\; C_K$ for certain structure functions $\kappa$ on 
the phase space. When the $\kappa$ are constant, the constraints generate a Lie algebra,
otherwise an algebroid. The case that there is an additional first class Hamiltonian $H$ 
with $\{C_I,H\}=\kappa_{I0}\;^K\; C_K$ can be embedded 
into the totally constrained case by adding the additional constraint $C_0=p_0+H$ and 
extending the phase space by the pair $(q^0,p_0)$. From here on the treatment splits 
into the constraint quantisation and reduced phase space quantisation route.

\subsubsection{Constraint quantisation}
\label{s2.1.1}

The starting point is construct a $^\ast-$algebra $\mathfrak{A}$. This is accomplished 
by selecting functions on the classical phase space which i. separate the points 
$(q,p)$ of the phase space, ii. close under complex conjugation and iii. close under Poisson brackets.
To avoid domain questions later, we will choose $\mathfrak{A}$ to be the Weyl algebra 
generated from Weyl elements $W(f,g)=\exp(i[f_a \;q^a+g^a\; p_a])$ where $f,g$ are 
certain real valued smearing functions $(f,g)$. 

The set of functions to which the $f,g$ belong has a strong impact 
for the set of states $\omega$ on $\mathfrak{A}$. Here a state is a positive ($\omega(a^\ast a)\ge 0$), 
linear ($\omega(z\;a+z'\;a')=z\omega(a)+z'\omega(a')$), 
normalised ($\omega(1)=1$) functional $\omega$ on $\mathfrak{A}$. Then the GNS construction 
provides GNS data $({\cal H},\rho,\Omega)$ consisting of a Hilbert space $\cal H$, a representation
$\rho$ of $\mathfrak{A}$ on $\cal H$ by operators and a normalised vector $\Omega$ cyclic 
for $\rho(\mathfrak{A})$. The set ${\cal D}=\rho(\mathfrak{A})\Omega$ is a common, dense and 
invariant domain for $\rho(\mathfrak{A})$ and we can compute scalar products via 
$<\rho(a)\Omega,\rho(b)\Omega>=\omega(a^\ast\;b)$. The representation is irreducible if and 
only if $\omega$ is pure. The unitary equivalence class of the GNS data is uniquely specified 
by $\omega$. We call a state $\omega$ regular with respect to $f$ if the 1-parameter unitary 
groups $t\mapsto \rho(W(t f,g))$ are strongly continuous, otherwise we call it irregular.
In the regular case we can define $\rho(f_a q^a)$ as a self-adjoint operator via 
Stone's theorem, in the irregular case we cannot. Similarly we define (ir)regularity with respect to $g$.   
If a state is regular with respect to both $f,g$ we simply call it regular.

In QFT the Stone von Neumann uniqueness theorem does not hold. Therefore among all possible states 
we must select physically admissible ones. The natural selection criterion consists in the 
requirement that the constraints can be promoted to at least quadratic forms $\rho(C_I)$ on 
$\cal H$. In the ideal case, they can be promoted to operators densely defined, preferably   
on the domain $\cal D$, in such a way that the domain is invariant. In that situation 
we can compute the commutator $[\rho(C_I),\rho(C_J)]$ on $\cal D$ and ask that it be a quantisation 
of $\kappa_{IJ}\;^K \; C_K$. In case that $\kappa$ is a constant function on phase space we 
therefore would require $[\rho(C_I),\rho(C_J)]=i\kappa_{IJ}\;^K\; \rho(C_K)$, if $\kappa$ 
is not constant then it becomes an operator by itself and the quantisation 
$\rho (\kappa_{IJ}\;^K \; C_K)$ suffers from operator ordering issues. For instance if 
$\rho(\kappa_{IJ}\;^K)$ can be defined independently, then an ordering compatible with 
a symmetric ordering of $\rho(C_I)$ is given by $\rho(\kappa_{IJ}\;^K \; C_K)=
\frac{1}{2}[\rho (\kappa_{IJ}\;^K)\;\rho(C_K)+\rho(C_K)\;\rho (\kappa_{IJ}\;^K)]$. 
If the $\rho(C_I)$ are merely quadratic forms densely defined on $\cal D$ then 
the formal image of $\rho(C_I){\cal D}$ typically lies outside of $\cal H$. That 
formal image is defined by $\rho(C_I)\psi:=\sum_n\; b_n\; <b_n,\rho(C_I)\psi>$ for 
$\psi\in {\cal D}$ and $b_n$ an orthonormal basis constructed from elements of 
$\cal D$ of $\cal H$. We allow that $\cal H$ 
be non separable, i.e. the indices $n$ could take values in an uncountable index set.  
In this situation, the commutator $[\rho(C_I),\rho(C_J)]$ is a priori ill-defined even as 
a quadratic form on $\cal D$. However, in fortunate cases it may still be defined 
using mode regularisation: we restrict the indices $n$ to a finite set $S_N$ controlled by 
some order parameter $N\in \mathbb{N}$ and replace $\rho(C_I)$ by $\rho(C_I)_N:=P_N\rho(C_I)P_N$
where $P_N$ is the orthogonal projection on the closure of the span of the 
$b_n,\;n\in S_N$. Then one computes $[\rho(C_I)_N,\rho(C_J)_N]$, takes $N\to \infty$ in the 
weak operator topology and shows that the limit exists and qualifies as a quadratic form 
quantisation of $\kappa_{IJ}\;^K C_K$ on $\cal D$. 

The vectors in $\cal H$ are not physical states. The physical states are generalised zero 
eigenvectors of the $\rho(C)_I$, i.e. linear forms $l$ on $\cal D$ such that 
$l[\rho(C_I)\psi]=0$ for all $I,\;\psi\in {\cal D}$. This equation is granted to 
be well defined when $\rho(C_I)$ preserves $\cal D$. But even if $\rho(C_I)$ is just a 
quadratic form, we can give meaning to this equation as follows:
Obviously $l$ is uniquely specified by the 
numbers $l_n:=l[b_n]$ and thus we may present $l$ as $l=\sum_n\; l_n\; <b_n,.>$. 
Then $l[\rho(C_I)\psi]=\sum_n\; l_n\; <b_n,\rho(C_I)\;\psi>$ and the matrix elements
$<b_n,\rho(C_I)\;\psi>$ are well defined by construction. Thus in both cases the physical state 
condition may be read as infinite system of linear equations for the coefficients $l_n$ and 
solutions $l_n^\ast$ will generically be generalised eigenvectors, i.e. $\sum_n \;l_n^\ast\; b_n$ is 
not a normalisable vector in $\cal H$. 

There is an important distinction between the case 
that $\cal D$ is an invariant domain for operators $\rho(C_I)$ and the case that the 
$\rho(C_I)$ are merely quadratic forms on $\cal D$. In the former case we trivially have 
$l[[\rho(C_I),\rho(C_J)]\psi]=0$ for all $I,J,\psi\in {\cal D}$ which implies
$l[[\rho(\kappa_{IJ}^K),\;\rho(C_K)]\psi]=0$ when all objects involved are defined as 
symmetric operators preserving $\cal D$. This implies conditions on $l$ that have no 
counter part in the classical theory indicating an {\it anomaly}. It is therefore conceivable that 
the $\rho(C_I)$, which are not observable and of which we want to know only the kernel, should not 
be symmetrically ordered. In case that $\rho(C_I)$ is not an operator with invariant domain $\cal D$ 
this implication does not hold because $l[\rho(C_I)\rho(C_J)\psi],\;l[\rho(C_J)\rho(C_I)\psi]$ 
do not exist separately, at best their difference does after regularisation as sketched above.

As the space of solutions $l$ typically consists of non normalisable states, we must equip it with 
a new Hilbert space structure. In case that $\kappa$ is a constant, in fortunate cases
we can pass from the Lie algebra $\mathfrak{g}$ generated by the $\rho(C_I)$ to the corresponding Lie 
group $G$ and define solutions $l$ by $\eta(\psi):=\int_G\; d\mu(g) <\rho(g)\psi,.>$ assuming 
that a Haar measure $\mu$ on $G$ exists. Then $<\eta(\psi),\eta(\psi')>_{{\sf phys}}:=\eta(\psi)[\psi']$ 
is a suitable inner product on the space of solutions. In case that these assumptions are not satisfied 
one must resort to different methods. One of them is the master constraint approach which replaces 
the classical set of conditions  $C_I=0$ by the equivalent single condition $M:=C_I\; Q^{IJ} \; C_J=0$
where $Q^{IJ}$ is a positive definite matrix valued function on the classical phase space. Then, if
$Q$ can be chosen such that 
$M$ is a self adjoint operator we can use the map $\eta$ where $G$ is the translation group. 

If the physical inner product can be obtained along those lines, the remaining task is to 
quantise Dirac observables on the resulting physical Hilbert space. Here a Dirac observable
is a function $O$ on phase space such that $\{C_I,O\}=\kappa_I^J(O)\; C_K$ where the
$\kappa(O)$ depend on $O$.\\ 
\\
Obviously, the constraint quantisation programme is quite involved and many steps can be faced with 
severe obstacles. This in part motivates the reduced phase space approach.

\subsubsection{Reduced phase space quantisation}
\label{s2.1.2}

In this route one considers as many gauge fixing conditions $G^I_t=x^I-\tau^I(t)=0$ as there are 
constraints $C_I=0$. The parameter $t$ plays the role of physical time.
We will not consider the most general case but assume from the outset 
that the $x^I$ are configuration functions on the phase space such that 
the matrix $\Delta_I^J:=\{C_I,x^J\}$ is non-degenerate. The $\tau^I(t)$ are constant functions
on the phase space that depend on a single parameter $t$. 
The condition on $\Delta$ grants that 
the gauge $G^I_t=0$ can be reached as $G^I_t\to G^I_t+u^I\; \Delta_J^I$ under a gauge transformation
with gauge parameter $u^I$ which can be solved for $u^I$. After the gauge has been installed
the gauge is preserved in time by the constrained Hamiltonian $H=u^I C_I$ if $u^I \Delta_I^J=\dot{\tau}^J$
which can be solved for $u^I=u^I_\ast(t)$. There are now two entirely equivalent
viewpoints: true degrees of freedom and relational Dirac observables. To establish this connection 
we split the canonical pairs $(q^a,p_a)$ into two sets $(x^I,y_I),(Q^A,P_A)$. The condition on $\Delta$
implies that we can solve $C_I=0$ for $y_I=-h_I(x,Q,P)$ using the implicit function theorem. 
We call $(x,y)$ the gauge degrees of freedom and $(Q,P)$ the true degrees of freedom. 
The latter are coordinates on the reduced phase space. We now 
restrict attention to functions $F$ on the reduced phase space i.e. which depend only on $(Q,P)$. Then the reduced 
Hamiltonian $E(t)$ is defined as that function on the reduced phase space which generates the same 
equations of motion as the constrained Hamiltonian $H=u^I\; C_I$ on the above gauge cut of the 
constraint surface , that is,
$\{E(t),F\}=\{H,F\}_{G=C=u-u_\ast=0}$. In general $E(t)$ will be explicitly time dependent.
An explicit formula fo $H(t)$ is obtained by passing from the $C_I(q,p)$ to the equivalent 
constraints $\hat{C}_I(q,p)=y_I+h_I(x;,Q,P)$. Then $E(t;Q,P))=\dot{\tau}^I(t)h_I(\tau(t),Q,P)$.
  
Equivalently, for any function $f$ on the unconstrained phase space coordinatised by $q,p$
we may construct the function $O_f(t):=[e^{X_u}\cdot f]_{u=-G_t}$ where $X_u$ is the Hamiltonian 
vector field of the function $u^I [\Delta^{-1}]_I^J \; C_J$. One can show that 
$\{C_I,O_F(t)\}=0$ when $C_J=0$  for all $J$ which means that $O_f(t)$ is 
a Dirac observable and that $\{O_f(t),O_g(t)\}=O_{\{f,g\}_\ast}(t)$
where $\{.,.\}_\ast$ is the Dirac bracket defined by $C_I,x^I$. 
%We will not treat the most general case 
%but confine ourselves to gauge fixings for which $\{G^I,G^J\}=0$ such as $G^I=x^I$ which has the advantage 
It follows that $\{F,G\}_\ast=\{F,G\}$ when $F,G$ depend only on $Q,P$. It is then not difficult to show 
that $\frac{d}{dt} O_F(t)=\{O_{E(t)},O_F(t)\}$ and $O_F(t)=F(O_Q(t),O_P(t))$. Thus
$(Q,P)\mapsto (O_Q(0),O_P(0))$ is a canonical transformation and the two viewpoints are equivalent.
We will pick the true degrees of freedom viewpoint for convenience but interpret the true degrees of freedom 
and reduced Hamiltonian gauge invariantly using the map $O_\cdot(t)$.

The reduced phase space quantisation now proceeds as in the previous subsection except that
we consider the reduced Weyl algebra $\mathfrak{A}$ generated by the $W(f,g)=\exp(i [f_A\;Q^A+g^A\; P_A])$. 
We select states on $\mathfrak{A}$ by requiring that $\rho(E(t))$ can be implemented as a self-adjoint 
operator. Obviously, the reduced phase space approach is simpler than the constraint quantisation approach 
as far as the quantum aspects are concerned. On the other hand, it maybe rather complicated to find $E(t)$ 
sufficiently explicitly. Moreover, to solve the system $C_I=0$ in the form $C_I=y_I+h_I=0$ may lead to several 
solution branches if the $C$ depend non-linearly on the $y$. Finally it maybe that also the solution 
$u^I_\ast$ has more than the 1-parameter ambiguity labelled by $t$ (Gribov copies).

\subsection{Application to CQG and LQG incarnation}
\label{s2.2}

In the first subsection we give a brief review of the canonical structure of classical GR including 
standard matter using two different polarisations (choice of canonically conjugate pairs) of the 
geometry part of the phase space. In the second we briefly review the constraint quantisation 
programme in the LQG incarnation.

\subsection{Polarisations}
\label{s2.2.1}

It turns out that GR perfectly fits into the general framework laid out in the previous subsection.
First of all we restrict to globally hyperbolic spacetimes $(M,g)$ such that the initial value problem 
is well posed. This implies that $M$ admits a foliation $t\mapsto \Sigma_t$ into spacelike hypersurfaces 
$\Sigma_t$ which are mutually diffeomorphic \cite{31,32} to some D-manifold $\sigma$. The foliation 
is now specified by a 1-parameter family of embeddings $Y_t: \sigma\to \Sigma_t$ displaying 
$M$ as diffeomorphic to $\mathbb{R}\times \sigma$. If $x^a,\; a=1,..,D$ denote coordinates on $\sigma$ 
then $(t,x)$ are known as Arnowitt-Deser-Misner coordinates. The timelike unit normal to the hypersurfaces 
can be parametrised as $n=\frac{1}{N}[\partial_t-N^a\partial_{x^a}]$ where $N,N^a$ are known as lapse scalar and 
shift vector field on $\sigma$.
  
In order to admit fermionic matter it is 
mandatory to require the existence of a spin structure and to introduce a (D+1)-Bein field $e_\mu^\alpha$ 
where both indices take range $0,..,D$, that is,
$g_{\mu\nu}=\eta_{\alpha\beta} e^\alpha_\mu e^\beta_\nu$ with Minkowski metric $\eta$. The gravitational Lagrangian can be 
chosen as the Einstein-Hilbert Lagrangian which just depends on $g$ but we must think of $g$ as 
composed of $e$ as above when fermions are present as the fermionic Lagrangian couples to the spin connection 
$\Gamma(e)$ of $e$ (so-called second order formalism). The canonical analysis of the resulting total Lagrangian including all matter 
and a cosmological term is well known, see e.g. \cite{5,20} and references therein. 
The phase space is totally constrained by first and second class constraints. The second class 
constraints can be solved by imposing the time gauge $e^0_\mu=n_\mu$ 
%where $e^\mu_\alpha$ is the inverse (D+1)-Bein 
which restricts the frame Lorentz group to the rotation group. Then the totally 
constrained Hamiltonian is a linear combination of constraints $H=N^a D_a+NC-\Gamma^{jk}_t\; G_{jk}+...$
where the dots stand for the various Gauss constraints of the Yang-Mills gauge groups of the standard model 
for matter. Here $G_{jk}$ is the gravitational Gauss constraint generating frame rotations. The constraints 
$D_a,C$ are called spatial diffeomorphism constraint (SDC) and Hamiltonian constraint (HC) respectively.

The constraints $G_{jk}=-G_{kj},\; j,k,l,..=1,..,D$ impose $D(D-1)/2$ conditions on the $D^2$ components 
of the D-Bein $e_a^j,\; a,b,c,..=1,..,D$ leaving $D(D+1)/2$ free components which coincides with the 
number of independent components of the D-metric $q_{ab}=\delta_{jk} e^j_a e^k_b$. There is a fermionic
contribution to $G_{jk}$ whose form depends on $D$ via the dimension dependence of the Clifford algebra
matrices $\gamma_\alpha$ and in general $D$ this contribution involve the rotation generator 
$[\gamma_j,\gamma_k]$.
In $D=3$ these contributions can be decomposed, after some relabelling, 
into several terms of the form $[b_A]^\ast [\tau_j,\tau_k]_{AB} b_B$
where $b^A,\;A=1,2$ are the components of a left chiral Weyl spinor and $\tau_j$ are 
basis elements of su(2) in the 2-dimensional representation normalised such that 
$[\tau_j,\tau_k]=\epsilon_{jkl}\tau_l$ (we use the Kronecker metric to raise and 
lower $j,k,l,..$ indices so that upper and lower case indices are indistinguishable). 
These Weyl spinors are related to the chiral 
components of the various Dirac
spinors that are present in the standard model and the fact that both $b,b^\ast$ appear which obey
canonical anti-commutation relations (CAR) $[b^A(x),[b^B]^\ast(y)]_+=\delta^{AB}\delta(x,y)$ 
allows us to write a right chiral spinor $b'$ contribution in terms of a left chiral one using 
the Majorana (charge conjugation) map $b'_A=\epsilon_{AB}\; [b^B]^\ast$ where $\epsilon$ is the 
spinor ``metric''. Specifically for $D=3$ and e.g. one spinor species
\be \label{2.1}
G_{jk}=e_{a[j}\; P^a_{k]}+i\; b^{\ast T}[\tau_j,\tau_k]\;b  
\ee
where $P^a_j$ is the momentum conjugate to $e_a^k$, i.e. the canonical commutation  relations 
read $[P^a_j(x),e_b^k(y)]=i\delta^a_b\delta_j^k\delta(x,y)$.    

The SDC is best written in its smeared version, e.g. in $D=3$
\be \label{2.2}
D[u]=\int_\sigma\; d^D\;[P^a_j \;[L_u e^j]_a+b^{\ast T} [L_u b]+ \pi\;[L_u\;\phi]+V^a\; [L_u\;W]_a]
\ee
for one fermion, one Higgs and one vector boson species $b,\phi,W$ respectively with respective conjugate 
momenta $b^\ast,\pi,V^a$. For more species (e.g. labelled by gauge group representation indices) 
simply sum over them. Here $L_u$ denotes the Lie derivative 
with respect to the vector field $u^a$ and is understood that $e_a^j,W_a$ are co-vector fields
of density weight zero, $b$ is a scalar of density weight $\frac{1}{2}$ and $\phi$ a scalar of
density weight zero. Of course there could be many more contributions coming from the hypothetical 
extensions (e.g. multi-Higgs), cosmological sector (e.g. inflaton, quintessence,..) or dark sector 
(e.g. axion) of the standard model. 

The HC is very complicated to write out in all details for all standard model contributions, however,
its essential structure can be described as follows
\be \label{2.3}
C=C_E+C_F+C_H+C_{YM}+C_I
\ee
where the individual contributions are referred to as Einstein, Fermion, Higgs, Yang-Mills and 
interaction pieces. We have again for $D=3$ 
\ba \label{2.4}
C_E &=& |\det(q)|^{-1/2}[P^{ab}\;P_{AB}-\frac{1}{D-1}P^a_a\;P^b_b]+|\det(q)|^{1/2}\;[2\Lambda- R[q]]
\nonumber\\
C_F &=& i\;b^{\ast T}\; e^a_j\; \tau^j\; [\nabla_a b]-c.c.
\nonumber\\
C_H &=& |\det(q)|^{-1/2}\frac{\pi^2}{2}+|\det(q)|^{1/2}\;[q^{ab}\frac{\phi_{,a} \phi_{,b}}{2}+U(\phi)]
\nonumber\\
C_{YM} &=& \frac{1}{4}[2\;|\det(q)|^{-1/2}\;q_{ab}\; \delta^{IJ}\;V^a_I V^b_J+
|\det(q)|^{1/2}\;q^{ab}\;q^{cd} {\sf Tr}(F_{ac} F_{bd})]
\ea
Here 
\be \label{2.4a}
\nabla\; b=(\partial+\Gamma(e)^j\;\tau_j+W^I X_I)\;b
\ee
is the gauge covariant derivative for fermions where $X_I$ are representation matrices of 
the Lie algebra basis of the Yang-Mills gauge group in question in the representation that 
$b$ is transforming in (e.g. the Gell-Mann matrices for SU(3)). $e^a_j$ denotes the inverse
D-Bein and $\Gamma_a^j(e)=-\Gamma_a^{kl}\epsilon_{jkl}/2$ is the spin connection of $e$. 
We have already reduced the phase space with respect to the isospin gauge invariance, thus 
$\phi$ is the real valued Higgs field with potential $U$ and the vector bosons $W_a^I,\;I=1,2,3$ 
are isospin gauge invariant (the isospin Gauss constraint is solved). We have not written 
out the interaction term $C_I$ which contains the Yukawa terms $\propto W \phi b$ for the weak 
interaction and mass terms due to a non-vanishing Higgs condensate. 
\be \label{2.5}
W_{ab}^I=2\partial_{[a} W_{b]}^I+W_a^J\;W_b^K\; f_{JK}\;^I,\;\;
[X_J,X_K]=f_{JK}\;^I \; X_I 
\ee
is the magnetic part of the Yang-Mills curvature while $V^a_I$ is the electric part.
Finally we abbreviated $P^{ab}=\frac{1}{2}\;P^{(a}_j e^{b)}_k\delta^{jk}$, all spatial 
tensor indices $a,b,c,..$ are raised and lowered with $q_{ab}=e_a^j e_b^k\delta_{jk}$, $\Lambda$
is the cosmological constant and 
$R[q]$ is the Ricci scalar of $q$.
We have refrained from writing out various coupling constants and Weinberg angle dependencies 
in order not to clutter the formulae.\\ 
\\
As already mentioned, a major role is played by the hypersurface deformation algebroid. 
Using smeared constraints $D[u], C[f]$ etc. a tedious computation reveals that 
\be \label{2.6}
\{D[u],D[v]\}=-D[L_u\;v],\;
\{D[u],C[f]\}=-D[L_u\;f],\;
\{C[f],Cg]\}=-D[q^{-1}[f\;dg-f\;df]
\ee
We also have the following relations with any of the smeared Gauss constraints $G[r]$
\be \label{2.7}
\{G[r],G[s]\}=-G[[r,s]],\;
\{D[u],G[r]\}=-G[L_u\;r],\;
\{C[f],G[r]]\}=0
\ee
revealing the first class nature of the constraint algebroid. Here $r,s$ are Lie algebra 
valued smearing fields and $[r,s]$ is their Lie algebra commutator.\\ 
\\
\\
This presentation of the classical canonical theory uses the $(e,P)$ polarisation of the 
classical phase space. It is available for any $D$. A special situation arises in    
$D=3$ \cite{12}. In that case the spin connection $\Gamma_a^j(e)$, which can also
be considered as a functional of $E^a_j:=|\det(e)| e^a_j$, can be written as 
$\Gamma_a^j=\frac{\delta S}{\delta E^a_j},\; S=\int_\sigma\; d^3x\; E^a_j\; \Gamma_a^j$.
This means that we have a canonical transformation $(e,P)\mapsto (A:=\Gamma+\gamma K,\;
\gamma^{-1} E)$ where $K_a^j=|\det(e)|^{-1}\; q_{ab} P^b_j$ to so called connection 
variables. The one parameter freedom $\gamma$ is called the Immirzi parameter. We will 
pick $\gamma=1$ in what follows to keep formulae simple. This means that in D=3 we 
also have a $(A,E)$ polarisation for the gravitational phase space similar to 
a Yang-Mills theory for the gauge group SU(2). In particular we can decompose 
$R=F+K\wedge K$ where $R$ is the Riemann curvature two form of $q$ in the triad frame and  
$F$ is the curvature of $A$. This fact has been historically the major motivation 
for considering this canonical transformation as it enables to discuss all interactions 
in the language of gauge theories of Yang-Mills type. If one wants to use connection variables 
in $D\not=3$ one must use the gauge group SO(D+1) instead of SO(3) and work with hybrid 
connections and additional simplicity constraints \cite{33}. 
 
\subsection{Constraint quantisation in LQG incarnation}
\label{s2.2.2}

The LQG incarnation uses the $(A,E)$ polarisation and therefore only is available in D=3. 
As connections do not form a vector 
space but rather an affine space, we have to slightly modify the canonical quantisation 
programme as outlined in the first subsection in terms of standard Weyl elements.
Motivated by similar (Wilson loop like) constructions in QCD we pick the following 
``generalised'' Weyl elements
\be \label{2.8}
W(f,g)=[{\cal P}\; e^{<f,A>}]\; e^{i<g,E>}  
\ee
Here the smearing fields are distributional ``form factors'' 
\be \label{2.9}
f^a(x)=\int_p dy^a\; \delta(x,y),\;
g_a^j(x)=\int_S\;\epsilon_{abc}\;dy^b\wedge dy^c \delta(x,y) g^j(y)
\ee
where $p,S$ are one and two dimensional submanifolds of $\sigma$ which come with 
some analyticity structure in order to control the Weyl relations. $\cal P$ is 
the path ordering symbol and $<f,A>=\int\;d^3x\; f^a A_a^j\tau_j,\;
<g,E>=\int\;d^3x\; g_a^j\; E^a_j$. Note that $W(f,g)$ is SU(2) valued and in order 
to obtain single operators we take one of its four matrix elements (the first factor 
in (2.9}) is called the holonomy of $A$ along $p$). The reason 
for this particular choice of Weyl elements is that they behave covariantly 
under spatial diffeomorphisms and Gauss rotations. We refer to the literature for 
the Weyl relations among the (\ref{2.8}) that result from the canonical 
conjugacy and reality of $(A,E)$. We denote the resulting $\ast$ algebra by 
$\mathfrak{A}$.

Let $\varphi$ be a spatial diffeomorphism. Then 
\be \label{2.10}
\alpha_\varphi(W(f,g))=W(\varphi\cdot f, \varphi\cdot g),\;
[\varphi\cdot f]^a= \int_{\varphi(p)} dy^a\; \delta(.,y),
[\varphi\cdot g]_a^j=\int_{\varphi(S)}\;\epsilon_{abc}\;dy^b\wedge dy^c \delta(x,y) g^j(\varphi^{-1}(y))
\ee
so that Diff$(\sigma)$ acts on $\mathfrak{A}$ by automorphisms. Then we have the following 
result \cite{34}.
\begin{Theorem} ~\\
Let $\omega=\omega\circ\alpha_\varphi$ be a spatially diffeomorphism invariant state 
on $\mathfrak{A}$ regular with respect to $g$. Then $\omega$ must be the 
LQG state.
\end{Theorem}
The LQG state \cite{35} can best be described in terms of spin network functions 
(SNWF) \cite{36}. These are products of matrix elements of irreducible SU(2) representations
(labelled by spins) evaluated on holonomies along piecewise analytic paths which 
are mutually disjoint except for their end points (a so-called graph).   
Any linear combination of products of matrix elements of holonomies can thus be
written as a finite linear combination of SNWF. Since by the Weyl relations any 
product of Weyl elements is such a polynomial in holonomy matrix elements times 
a factor of the $e^{i<g,E>}$ it is sufficient to specify the state on elements 
of the form $T \;e^{i<g,E>}$ where $T$ is a SNWF. We then have 
\be \label{2.11}
\omega_{{\sf LQG}}(T \;e^{i<g,E>})=\delta_{T,1}
\ee 
i.e. (\ref{2.11}) vanishes unless $T$ is the trivial SNWF $T=1$. One can show 
that this implies that SNWF form an orthonormal basis. In fact such type of states 
were discovered earlier for different Weyl algebras in the context of QED \cite{360}.
We draw the following
conclusions from this observation.
\begin{itemize}
\item[0.] The construction of $\mathfrak{A},\omega$ is manifestly background independent.
The GNS Hilbert space has a Schr\"odinger like realisation as an $L_2$ space of square 
integrable functions with respect to a probability measure on a space of distributional 
connections \cite{36a}.  

\item[1.] Since the set of graphs
is uncountably infinite, the resulting GNS Hilbert space is not separable.     
\item[2.] Since at most a countable superposition of SNWF can be normalisable, 
normalisable semiclassical states saturating the Heisenberg uncertainty bound for 
Weyl elements \cite{36b}, which we think of as superpositions of excitations of the GNS vacuum
$\Omega$, can only depend on a countable union of finite graphs and thus a 
countable number of vertices. While such a vertex set can be dense in $\sigma$ \cite{24}
the set of  
points in $\sigma$ at which the quantum volume (see below) vanishes is has full Lebesgue 
measure. 
\item[3.] There is no operator valued distribution corresponding to $A$. In 
order to define $A$ one would consider a 1-parameter family of paths $p_s(t)=p(st),\; 
s,t\in [0,1]$ and then would try to define \\
$\dot{p}^a(0) <T,A_a^j(p(0))T'>:=
-2[\frac{d}{ds}{\sf Tr}(\tau_j\;<T,W(f,0)T'>)]_{s=0}$ but this trivially vanishes because 
$\omega$ is irregular with respect to $f$.
\item[4.] Diff$(\sigma)$ is represented unitarily by $U(\varphi)T=\alpha_\varphi(T)$. 
\item[5.] There is no operator corresponding to $D[u]$. In order to define 
$D[u]$ one would consider the 1-parameter family of diffeomorphisms $\varphi^u_s(x)=c^u_x(s)$ defined 
by the integral curves of $u$ and then would try to define 
$i\;<T,D[u] T'>=\frac{d}{ds}<T,U(\varphi^u_s)T'>]_{s=0}$ but again the right hand side trivially 
vanishes.
\item[6.] Since holonomies transform as $[{\cal P}\; e^{<f,A>}]\mapsto g(p(0))\;
[{\cal P}\; e^{<f,A>}]\;g(p(1))^{-1}$ under local gauge SU(2) transformations, 
the local gauge group acts by automorphisms $\beta_g$ and $\omega$ is also invariant under those. 
Accordingly $U(g)T=\beta_g(T)$ is a unitary representation which is in fact strongly 
continuous and therefore a Gauss constraint operator $i\;G[r]\;T=
[\frac{d}{ds} U(e^{sr^j\tau_j})T]_{s=0}$ exists. 
\item[7.] It is easy to find the kernel of the Gauss constraint. Without fermions these are simply 
the SNWF over closed graphs (no endpoints) such that for each vertex (necessarily 
at least bi-valent) all spin representations corresponding to adjacent edges are intertwined to 
the trivial representation. Thus solutions to the Gauss constraint remain normalisable 
in the GNS Hilbert space. Similar statements hold when one invokes fermions \cite{36c}
\item[8.] Solutions to the SDC are no longer normalisable \cite{37}. Leaving out some 
details about graph symmetries, we may 
define them as linear forms $[T]=\sum_{T'\in {\sf orb}(T)}\; <T',.>$ where 
orb$(T)=\{U(\varphi)T;\;\varphi\in{\sf Diff}(\sigma)\}$ is the orbit of $T$ under 
diffeomorphisms. It follows $[T](T')\propto\chi_{{\sf orb}(T)}(T')\in \{0,1\}$ ($\chi_S$ is 
the characteristic function of the set $S$). It is precisely due to the irregularity of 
$\omega$ that these linear functionals are well defined. Moreover an infinite 
number of inner products on the span of the $[T]$ maybe defined via 
$([T],[T'])=c_{[T]} \;[T'](T)$ where $c_{[T]}>0$ is arbitrary. 
\item[9.] There exist length \cite{38}, area \cite{39} and volume \cite{40} operators 
$L[c],A[S],V[R]$ for 
curves, surfaces and regions $c,S,R$ which have pure point spectrum. This happens precisely 
because $W(f,g)$ is regular with respect to $g$ and the spectral discreteness is due 
to the fact that SU(2) is compact (basically $E^a_j$ acts as an angular momentum operator).
\item[10.] As far as the HC is concerned, we face the following problem. Since
$C$ depends on the curvature $F$ of $A$ which is a second order polynomial in $A$, but 
$A$ does not exist in this representation, we must regulate $C$ by approximating 
$F$ by a one parameter family of holonomies along closed loops $p_s$ of coordinate 
size $s^2$, divide by $s^2$ and then take $s\to 0$. It turns out that if and only if 
$C$ carries density weight unity can $s$ be absorbed into a regularisation of an 
inverse volume operator and the limit becomes trivial in an operator topology 
of weak $^\ast$ type involving spatially diffeomorphism invariant states \cite{41}. Thus 
the HC can be defined {\it non-perturbatively and background independently} 
on the GNS Hilbert space as an operator with the span ${\cal D}$ of SNWF as 
dense invariant domain but the limit is ambiguous because 
it depends, among other things, on the diffeomorphism class of the loop family and 
the linear combination of SU(2) representation on the loop holonomy in terms of 
which it is regularised. This in fact only quantises the $F$ dependent term in $C$
which is called the Euclidian contribution $C_E$. However, one can now use the classical 
identity $K=\{V[\sigma],C_E[1]\}$ to define also the $K$ dependent term.
\item[11.] On the 
other hand, the GNS vacuum $\Omega$ is a ground state for the HC i.e. $C[f]\Omega=0$ 
for all $f$ which makes the LQG state a ``dynamically chosen'' state. This, together with the fact that 
the smearing dimension 1 and 2 respectively of $A$ and $E$ respectively is dual and adds
to $D=3$ is responsible for the fact $C[f]$ can be defined as an operator on $\cal D$. 
\item[12.] As far as the HDA is concerned, it can certainly not be implemented as 
in (\ref{2.6}) with Poisson brackets replaced by commutators because we do not have $D[u]$
at our disposal. The first two relations in (\ref{2.6}) can be dealt with by reformulating them as 
$U(\varphi)U(\varphi')=U(\varphi\circ \varphi'),\; U(\varphi)\;C[f]\; U(\varphi)^{-1}=C[f\circ\varphi])$
but the third relation cannot because of the presence of $q^{-1}$. The problem that 
$q^{-1}$ appears to diverge at points of $\sigma$ which are not vertices of the underlying 
graph can be dealt with \cite{41}. At least for vacuum 
gravity it is possible to show that $l[[C[f],C[g]]\;T]=0$ for any SNWF $T$ and lapse functions $f,g$ 
when $l$ is a solution of the SDC. This means that $[C[f],C[g]]T$ is of the form $\sum_{k=1}^N (U(\varphi_k-1_{{\cal H}})\;O_k\;T$
for some $N<\infty$ where $\varphi_k$ and the operators $O_k$ depend on $f,g,T$. Hence the 
HC close up to a linear combination of SDC but the $O_k$ turn out not to qualify as quantisations
of $q^{-1}$ (wrong structure functions). In that sense there is no mathematical inconsistency
but still a physical anomaly.
\end{itemize}
Several proposals have been made to improve the situation described in items 
[10.], [12.]. One can try to formulate $C$ on some dual of the span $\cal D$ of SNWF
\cite{42}, one can replace the infinite number of $C[f]$ by the single master constraint
$M\int\;d^3x\;|\det(e)|^{-1} C^2$ which trivially obeys $[M,M]=0$ and of which we just 
need to require $[U(\varphi),M]=0$ \cite{25} or we can work on the reduced phase 
therefore solving and implementing the HDA classically using suitable matter \cite{21}
along lines of the previous section. While these proposals turn
out to be at least partly successful, all of them suffer from the ambiguity issue described 
in item [10.]. Renormalisation methods could help to remove part of these ambiguities \cite{22}.
However, as all of these problems find their roots in the choice of $\mathfrak{A},\omega$
we ask in the next section for alternative, background independent and non-perturbative 
possibilities.         
     
\section{Background independent, non-perturbative Fock incarnation of CQG}
\label{s3}

The first subsection is devoted to a detailed analysis of the classical symplectic geometry 
of the geometry and matter content of GR plus standard model and their interplay with 
the various Gauss, the spatial diffeomorphism and the Hamiltonian constraints of the theory.
Joined with the desire to keep versions of these constraints polynomial and to construct 
scalar densities of weight $\frac{1}{2}$ that transform covariantly under Gauss and SDC 
in order to keep background independence dictate much of the structure of the available canonical
variables. In the second subsection we then introduce the background independent Fock representations 
based on those variables.

\subsection{Classical preparation}
\label{s3.1}  

In the first subsection we derive the necessity to work with scalar densities of weight
$1/2$ and why the construction of such canonical variables that stay purely within 
the geometry sector is not possible. After that we investigate possible solutions to this 
problem when combining geometry and matter variables while keeping gravitational Gauss covariance. 
Next we show that those solutions that in fact solve the gravitational Gauss constraint are 
singled out in the sense that in this case we are able to keep also the (rescaled) Hamiltonian 
constraint simple. Then we briefly show that the analysis drastically simplifies when one 
additionally solves the the SDC and the HC, details of which will be subject of another
manuscript.  

\subsubsection{Motivation}
\label{s3.1.1}

As the LQG incarnation shows, while the distributional smearing is very natural from the 
point of view of i. background independence ($A,\ast E$ are 1-forms and 2-forms respectively),
ii. diffeomorphism and Gauss covariance and iii. $C[f]$ being an operator in density weight 1 rather 
than merely a quadratic form, it is problematic from the point of view of having quantum states 
which are non-degenerate as one in fact silently assumes in the classical theory. 
Next, while holonomies are natural objects to consider when trying to solve a Gauss constraint,
as far as the HC is concerned an $(A,E)$ polarisation does not simplify 
the HC too much, especially not in density weight 1, as compared to a $(e,P)$ polarisation. 
This is because of the $K\;K$ term which would be missing for the original self-dual choice $\gamma=i$
\cite{12} but such a polarisation suffers from very complicated reality (adjointness) conditions which 
so far could not be rigorously implemented, see \cite{43} for some proposals. Therefore we stick to the original 
$(e,P)$ polarisation which also has the (debatable) advantage of being available in any dimension
$D$. 

We are therefore motivated to consider the standard Weyl elements for the gravitational sector 
\be \label{3.1}
W(f,g):=e^{i[<f,e>+<g,P>]},\; <f,e>=\int_\sigma\;d^Dx\; f^a_j\; e_a^j,\;
<g,P>=\int_\sigma\;d^Dx\; g_a^j\; P^a_j
\ee
As the space of D-Bein fields and their conjugate momenta forms a vector space (we drop
the requirement of non-degeneracy of $e$ at this point) this object is geometrically meaningful. 
We also have a covariant action of spatial diffeomorphisms $\varphi$ and gravitational Gauss 
rotations $O$ 
by automorphisms 
\be \label{3.2}
\alpha_\varphi(W(f,g))=W([\varphi^{-1}]^\ast f,[\varphi^{-1}]^\ast g),\;
\beta_O(W(f,g))=W(O^{-1} f,O^{-1} g),\;
\ee
where $\varphi^\ast$ is the tensorial pull-back and $f$ is considered as a vector density of weight one 
while $g$ is considered as a co-vector of weight zero. Both $f,g$ transform in the defining representation
of SO(D). 

We now ask for an invariant state $\omega=\omega\circ\alpha_\varphi=\omega\circ\beta_O$ which therefore is
automatically background independent and leads to unitary representations of both the spatial diffeomorphism and 
local Gauss gauge group.
A possible choice is the state of Narnofer-Thirring type
\be \label{3.3}
\omega(W(f,g))=\delta_{f,0}\delta_{g,0}
\ee
but this state is irregular and leads again to a non-separable Hilbert space with all the complications 
of the LQG state mentioned in the previous section. 
Therefore we add the requirement of regularity with respect to both $f,g$. 

Examples of regular states on the Weyl algebra generated by the $W(f,g)$ are Fock states
\be \label{3.4}
\omega(W(f,g))=\exp(-\frac{1}{4}[<f,\kappa^{-2}\cdot f>_1+<g,\kappa^{2}\;g>_1)  
\ee
%a=(k q-i p/k)/s, s=\sqrt{2}, fq+gp=f/k[a+a*]s+igk(a-a*)/s=(f/k+igk)/s a+(f/k-igk)/s a*=za+\bar[z} a*
% e^A e^B=e^{A+B} e^{[A,B]/2}
%exp(i[fq+gp])=\exp(i\bar[z}a*)\exp(iza)\exp(-[i\bar{z}a*,iza]/2)=(.)* (.) \exp(-|z|^2/2)=(.)* (.) \exp(-[f^2/k^2+g^2 k^2]/4)
where $\kappa(x,y)=\kappa(y,x)$ is a positive (thus symmetric) invertible integral kernel on 
the one particle Hilbert space $\mathfrak{h}_1$ with one particle inner product $<f,f>_1=\delta_{ab}\delta^{jk}
\int_\sigma\;d^Dx f^a_j f^b_k$. However, this construction lies at the other extreme of being 
heavily background dependent (here using the flat backround metric $\delta_{ab}$). The corresponding 
annihilation operator is 
\be \label{3.5}
a_b^j=([\kappa\cdot e]_a^j-i\delta_{ab}\delta^{jk} [\kappa^{-1}\cdot P]^b_k]/\sqrt{2}
\ee
and the GNS Hilbert space is the $L_2$ space with respect to a Gaussian measure
supported on the Schwartz distributions with covariance $\kappa^{-2}$.   

From a tensorial point of view, the annihilator is geometrical nonsense. It is a linear combination 
of tensor fields of different types (different density weight and different number of contra-variant and 
co-variant indices). It is therefore impossible to construct BI Fock representations unless one finds
BI substitutes of $\kappa,\delta_{ab}$. More promising are BI annihilation operators of the form 
\be \label{3.5a}
a_b^j=(\frac{\delta S}{\delta e_a^j}-i P^a_j)/\sqrt{2}
\ee      
for positive (semidefinite), spatially diffeomorphism and Gauss invariant functionals $S$ of $e$ such as 
the total volume $S=\int\;d^D\; |\det(e)|$ but the formal Hilbert space based on 
the vacuum state $\Omega=e^{-S}$ in a formal $L_2$ space
with respect to the functional Lebesgue measure $[de]=\prod_{x,a,j} de_a^j(x)$ 
is very hard to define and certainly does not correspond to a Fock state as the commutator
$[a,a^\ast]$ is not a constant times the unit operator. 

A hint for how to proceed stems from the fermionic sector. There we simply define the Fock vacuum
by the condition $b\Omega_F=0$. The normal ordered SDC for fermions is a densely and invariantly defined on the corresponding 
span of Fock states as $:D[u]:=\int \; d^Dx\; \frac{u^a}{2}[b^\ast b_{,a}-b^\ast_{,a}\; b]$ contains no 
$(b^\ast)^2$ terms. Tracing back from where this remarkable property originates one finds that it is the
fact that $b$ is a {\it scalar density of weight $\frac{1}{2}$}. Indeed, if one would be able to construct 
a complex linear combination from $e,P$ such that the resulting object is a complex valued tensor of definite 
type and weight then an annihilation operator maps to an annihilation operator under diffeomorphisms. Otherwise it maps 
to a linear combination of annihilation and creation operators (non-trivial Bogolubov transformation) 
and thus the Fock vacuum cannot be diffeomorphism invariant. More information can be found in \cite{16}. 
    
Thus in order to define a background independent Fock representation for the gravitational or matter fields 
we must assemble a complex linear combination $a$ of $e,P$ such that $a$ is a tensor density of some weight.
Denoting $A-$times contravariant, $B-$times covariant densities of weight $w$ by $(A,B,w)$ we find 
that $e,P$ are respectively of dual types $(0,1,0)$ and $(1,0,1)$ respectively. As we want the commutator 
$[a,a^\ast]$ to be a constant times the identity operator, the most general Ansatz for $a$ that we can 
make and that stays within the pure geometry sector would be 
\be \label{3.5b}
a^a_j=\frac{1}{\sqrt{2}}[((I\cdot e)^a_j-i\;P^a_j]
\ee
where $I$ is a phase space independent, positive, invertible intertwiner between the $(0,1,0)$ 
and $(1,0,1)$ tensors. A natural choice would be $(I\cdot e)^a_j=\delta_{jk} \; \epsilon^{abc}\partial_b e_c^k$
however this intertwiner fails to be invertible (it kills exact 1-forms) and positive. Moreover it destroys 
another desirable property namely that $a$ transforms in the defining representation of SO(3). The latter
could be achieved by replacing $\partial$ by $\nabla=\partial+\Gamma$ where $\Gamma$ is the spin connection 
but by construction $\nabla e=0$. While we do not have a conclusive proof it appears that a BI Fock state 
that stays within the pure geometry sector does not exist. 

\subsubsection{Canonical matter-geometry variables}
\label{s3.1.2}

In \cite{16} it was observed that BI Fock states exist when one is willing to mix geometry and matter sector.
Here we extend this observation. Before going into details, let us explain the basic idea:\\
Prior to quantisation we perform canonical transformations on the combined geometry and matter phase space.
These are split into three steps. The first step consists in producing D scalar fields $X^I$ of weight zero with 
conjugate momenta $Y_I$ which are scalar fields of weight 1 via an explicitly invertible 
canonical transformation out of 
the available geometry and matter field. These D scalar fields could already be present depending on how large 
the scalar particle, dark and cosmological sector are. If not, it turns out that one can assemble them, 
at least in $D=3$, from the weak vector bosons. In the second step, we consider the flat D-Bein $h_a^I=X^I_{,a}$.
Assuming that $\det(h)\not=0$ we can use it to turn all tensor fields, except $X^I,Y_I$  into scalar densities of weight $\frac{1}{2}$ by 
an explicitly invertible canonical transformation. 
%In particular we now have access to the density $\frac{1}{2}$ scalars $\hat{e}_I^j$. 
In the third step we map also $X^I,Y_I$ to density $\frac{1}{2}$ scalars using the now scalar density $\frac{1}{2}$ valued 
geometrical D-bein $\hat{e}_I^j$ via an 
explicitly invertile canonical transformation.
 
These transformations must maintain manifest covariance with respect to the 
gravitational Gauss and the spatial diffeomorphism constraint. This forces us to solve the gravitational 
Gauss constraint via symplectic reduction along the way if we use the route via the weak vector bosons because the 
scalar densities $X^I,Y_I$ should be Gauss invariant. In order to write the Hamiltonian constraint as a polynomial 
after multiplying it with a single phase space dependent factor, a very specific choice of Gauss reduced variables
is singled out which is related to the upper triangular gauge \cite{24}. If such Gauss invariant scalar densities are already 
present we are not forced to solve the gravitational Gauss along the way. However, again if a polynomial version 
of the Hamiltonian constraint is desired, the intermediate Gauss reduction appears to be mandatory, again using the techniques 
of \cite{24}. We will show later how to define certain versions of non-polynomial Hamiltonian constraints as quadratic 
forms but of course in this version the commutator of Hamiltonian constraints is even more complicated to 
compute. 
    
As the end result now consists in a phase space 
description in which all bosonic and fermionic fields are canonical pairs of scalar densities of weight $\frac{1}{2}$ we can 
take complex linear combinations of those with constant coefficients to construct covariant annihilation operators, define 
background independent and non-perturbative Fock representations and eventually SDC and Gauss constraint as self-adjoint 
operators.
\\
The details are as follows:
\begin{itemize}
\item[1.] {\bf Step 1:}\\
For completeness we consider the case that the D scalar fields are not already available but that the matter sector 
contains D canonical pairs $(W_a^I, V^a_I),\;I=1,..,D$ where the index $I$ does not represent a representation
label for some gauge group but rather labels those pairs such as the weak vector boson and its conjugate 
momentum after reducing with respect to the isospin gauge group (Higgs mechanism). Here $W,V$ are ``born'' 
in the tensor type 
$(0,1,0)$ and $(1,0,1)$ respectively by the canonical (Hamiltonian) formulation. We now define 
\be \label{3.6}
\tilde{W}^I_j:=W_a^I\; e^a_j,\; \tilde{V}_I^j:=V^a_I\; e_a^j,\; 
\tilde{e}_a^j:=e_a^j,\; \tilde{P}^a_j:=P^a_j+W_b^I \; V^a_I\; e_b^j  
\ee
%V^a_I d(W_j^I e^j_a)=V^a_I W^I_j de_a^jj
It is easy to check that $<P,\delta e>+<V,\delta W>-[<\tilde{P},\delta \tilde{e}>+<\tilde{V},\delta \tilde{W}>]$
is exact, hence the transformation is canonical. It has the inverse 
\be \label{3.7}
W^I_a=\tilde{W}_j^I\; e^j_a,\; V_I^a=\tilde{V}_I^j\; e^a_j,\; 
e_a^j=\tilde{e}_a^j,\; P^a_j=\tilde{P}^a_j-\tilde{W}_j^I \; \tilde{V}^k_I\; \tilde{e}^a_k  
\ee
The transformations (\ref{3.6}), (\ref{3.7}) rely on the assumption that $\det(e)\not=0$.
The subset of D-Beine such that $\det(e(x))=0$ defines a complicated submanifold $M(x)$ in $\mathbb{R}^{D^2}$ 
of co-dimension one depending on $x\in\sigma$ with several branches defined by the roots of a homogeneous polynomial of degree 
$D$ in $D^2$ real variables. Thus, for each $x$, $M(x)$ has zero measure with respect to the 
Lebesgue measure on $\mathbb{R}^{D^2}$. In classical GR one strictly speaking only allows 
for D-metrics with $\det(q)=\det(e)^2>0$ and one approach is to quantise the cotangent 
bundle over the non-degenerate D-Bein configuration space. However, as this is no longer 
a vector space, we will therefore ignore those degenerate configurations in the quantisation 
process and consider non-degeneracy as a semi-classical concept which however can be lifted 
for generic quantum states. We will come back to this issue later.

We now write the gravitational Gauss and SDC in the new variables (we only consider the 
contributions to the constraints that 
are affected by the canonical transformation)
\ba \label{3.8}  
D[u] &=& \int\; d^Dx\;(P^a_j\; [L_u e^j]_a+V^a_I\;[L_u W^I]_a)
=\int\; d^Dx\;(\tilde{P}^a_j\; [L_u \tilde{e}^j]_a+\tilde{V}^j_I\;[L_u \tilde{W}^I_j])
\nonumber\\
G[r] &=&  \int\; d^Dx\; r^{jk}\; P^a_j e_{ak}
=\int\; d^Dx\; r^{jk}\; (\tilde{P}^a_j \tilde{e}_{ak}+\tilde{V}_{Ij} \tilde{W}^I_k)
\ea 
where $r^{jk}=-r^{kj}$ is so(D) valued. Note that the SDC now displays $\tilde{V}^j_I,\tilde{W}_j^I$ 
as tensors of type $(0,0,1)$ and $(0,0,0)$ respectively. The price to pay is that these 
now transform in the defining representation of SO(D). 
\item[2.] {\bf Step 2:}\\
As we want to obtain not only spatially diffeomorphism covariant annihilators but also 
Gauss co-variant ones, we cannot simply use $D$ of the $D^2$ fields $W^I_j$ to construct 
a flat D-Bein. We must instead use $D$ of the $D(D+1)/2$ SO(D) invariant scalars 
$m^{IJ}:=\tilde{W}^I_j\tilde{W}^J_k \delta^{jk}$ which defines a positive symmetric matrix. 
We can conveniently perform this as 
follows, using in fact the idea of the Higgs mechanism:

Recall that for the Higgs mechanism we start with a complex iso-dublett $\Phi=(\Phi_1,\Phi_2)^T$ which 
transforms in the two dimensional (defining) representation of SU(2), $\Phi\mapsto g\Phi$. From this we construct 
the SU(2) matrix $U=(\Phi,-\epsilon \Phi^\ast)||\Phi||^{-1}$ which relies on the assumption 
that $\phi^2:=||\Phi||^2=|\Phi_1|^2+|\Phi_2|^2>0$. Here $\epsilon=i\sigma_2$ (Pauli matrix) is the 
totally skew symbol in 2 dimensions. Note that $<-\epsilon\Phi^\ast,\Phi>=0,\;||-\epsilon\Phi^\ast||=||\Phi||$
so $\Phi,-\epsilon\Phi$ is an orthogonal basis in $\mathbb{C}^2$ with respect to this scalar product 
on $\mathbb{C}^2$ and thus 
$\det(U)=1$. By construction $U\mapsto g\;U$. We then use $U$ to 
turn the gauge dependent W-bosons into gauge invariant ones (3 Higgs degrees of freedom have 
become Goldstone bosons). The only remnant of $\Phi$ in the 
Hamiltonian is the real valued Higgs $\phi$ and the iso-Gauss constraint can be explicitly 
and algebraically solved.

Here we proceed completely analogously: The role of $U$ is played by an SO(D) matrix $O$
constructed as follows. In fact, there are many ways of doing this and we display 
two relatively simple constructions:\\
\\
A. {\bf Polar decomposition}\\ 
We assume that the $D$ vectors $\tilde{W}^I$ with D entries $\tilde{W}^I_j$ are linearly independent, i.e.
$|\det(\tilde{W})|>0$. This condition is the analog of $||\Phi||^2>0$ above. Consider the 
positive real valued symmetric SO(D) invariant matrix $m^{IJ}:=W^I_j W^J_k\delta^{jk}$. It has 
a real valued {\it symmetric square root} $S^{IJ}$, i.e. $m^{IJ}=S^{IK}\;\delta_{KL} \; S^{LJ},\; S^{IJ}=S^{JI}$ which 
can be chosen to be positive but we will not require this. To see this we use that $m^{IJ}$ has 
real eigenvectors $B^\alpha_I,\;\alpha=1,..,D$ with positive eigenvalues $z_\alpha$ i.e.
$[m^{IJ}-z_\alpha \; \delta^{IJ}]B^\alpha_j=0$ which are 
orthogonal $B^\alpha_I B^\beta_J \delta^{IJ}=\delta^{\alpha\beta}$. 
Then pick $r_\alpha=\epsilon_\alpha\sqrt{z_\alpha}$ and $S^{IJ}=\delta^{IK} \delta^{JL}\sum_\alpha 
r_\alpha B^\alpha_K B^\alpha_L$ with $\epsilon_\alpha =\pm 1$. Let now $O_{Ij}:=S^{-1}_{IJ} W^J_j$ then $O_{Ij} O_{Jk} \delta^{jk}=\delta_{IJ}$ 
hence $O_{Ij}$ is orthogonal and we may pick the sign of $\det(S)$ to equal the sign of $\det(W)$ so that 
$\det(O)=1$ i.e. $O\in$ SO(D) by exploiting the freedom in picking $\epsilon_\alpha$. 
Finally note that $S^{IJ}$ is constructed from SO(D) invariants. 
Thus the vectors $O_{I}$ with entries $O_{Ij}$ transform in the defining representation of SO(D)
as the vectors $W^I$ with entries $W^I_j$. Accordingly we have $W^I_j=S^{IJ}\; O_{Jj}$ which is a generalised version 
of the polar decomposition as we can ensure that $\det(O)=1$ at the price of having $S$ indefinite. Note 
that to construct $S^{IJ}$ explicitly we must use the spectral theorem, hence the characteristic 
polynomial, for which we can write its roots no longer in closed form for $D>4$ but fortunately we do not need
them.\\ 
\\
B. {\bf Upper triangular decomposition}\\
Here we follow \cite{24} and pick an {\it upper triangular square root} $U^{IJ}$  
i.e. $m^{IJ}=U^{KI}\;\delta_{KL} \; U^{LJ};\; U^{IJ}=0$ for $I>J$ ($I,J$ considered as 
row respectively column index) which 
can be chosen to have positive diagonal entries but we will not require this.  
One of the advantages of this choice is that this system of quadratic equations can be solved 
in $D(D+1)$ steps in closed form for any $D$. We have $m^{11}=[U^{11}]^2$ hence $U^{11}=\epsilon_1\sqrt{m^{11}}$.
Then $m^{J1}=U^{11}\; U^{1J}$ hence $U^{1J}=\frac{m^{J1}}{U^{11}},\;J>1$. Next 
$m^{22}=[U^{12}]^2+[U^{22}]^2$ hence $U^{22}=\epsilon_2\sqrt{m^{22}-[U^{12}]^2}$.  
Then $m^{J2}=U^{12}\; U^{1J}+U^{22}\;U^{2J}$ hence $U^{2J}=\frac{m^{J2}-U^{12}\; U^{1J}}{U^{22}},\; J>2$ etc.
Then we define $W^I_j=:U^{JI}\; O_{Jj}$, verify that $O_{Jj}$ is orthogonal and use the freedom in the 
choice of $\epsilon_I$ to ensure that $O$ is unimodular at the price of having indefinite diagonal entries 
$U^{II}$. Note the important property that $\det(U)=\prod_{I=1}^D\; U^{II}$. By construction, the $U^{IJ}$ 
are SO(D) invariant scalars. \\    
\\
For both decompositions we now use the assumed spin structure in order to construct a matrix $C$ in the covering group of SO(D)
whose representation matrix in the $D$ dimensional representation coincides with $O$. 
E.g. in $D=3$ we must solve the equation $R_{Ij}(C):=-2{\rm Tr}(C\tau_j C{^\ast T}\tau_I)=O_{Ij}$ for $C\in$ SU(2)
which is invariant under $C\mapsto -C$.
It is precisely 
the purpose of a spin structure to ensure that one can solve consistently over the atlas of $\sigma$ 
the arising sign ambiguity. Denote by $C(O)$ such a consistent solution. Then by construction 
$C(g O)=C(g)\;C(O)$ for any $g\in$ SO(D).

Given $O$ constructed from $W$ we may now construct the SO(D) invariant spinors $\hat{b}=C(O)^{-1} b$ and 
the SO(D) invariant D-bein $\hat{e}_a^I:=\tilde{e}_a^j \delta^{IJ} O_{Jj}$. Then by construction 
the spinor covariant derivative is equivariant and we have the identity (here for $D=3$) 
$b^{\ast T} e^a_j \tau^j \tilde{\nabla}_a b=\hat{b}^{\ast T} \hat{e}^a_I \tau^Ij \hat{\nabla}_a \hat{b}$ where 
$\hat{\nabla}b=\partial b+\hat{\Gamma}^I\tau_I$ were $\hat{\Gamma}^I$ is constructed from $\hat{e}$ in 
exactly the same way as $\tilde{\Gamma}^j$ is from $\tilde{e}$. 

We now perform another canonical transformation.\\
\\
{\bf Polar decomposition}\\
\\
 For the polar decomposition   
\be \label{3.9}
\hat{e}_a^I:=\tilde{e}_a^j\; \delta^{IJ}\;O_{Jj},\;\hat{P}^a_I:=\tilde{P}^a_j\; O_{Ik}\delta^{jk},\;
\tilde{W}^I_j=:S^{IJ} O_{Jj},\; V^+_{IJ}:=\tilde{V}_{(I}^j O_{J)j)},\;
V^-_{IJ}:=[\tilde{P}^a_j \tilde{e}_{ak}+\tilde{V}_{Lj} W^L_k]\; M^{jk}_{IJ}
\ee
where where $S,V^+$ are symmetric, $V^-$ is antisymmetric, $O$ is both orthogonal 
and unimodular as constructed above and
\be \label{3.10}
M^{jk}_{IJ}:=\delta^{KL} \delta^{jm}\;\delta^{kn}\; O_{Km}\; \frac{\partial O{Ln}}{\partial \theta^{IJ}}
\ee
Here $\theta^{IJ}=-\theta^{JI}$ are $D(D-1)/2$ Euler angles parametrising $O$
and (\ref{3.10}) is to be evaluated at $\theta=\theta(O)$. By construction,
$M^{jk}_{IJ}$ is separately antisymmetric in both index pairs and non-degenerate on the space 
of anti-symmetric matrices. 
It is easy to check that the difference between the symplectic potentials is exact where 
$V^+_{IJ}$ is conjugate to $S^{IJ}$ and $V^-_{IJ}$ to $\theta^{IJ}$.
The inverse transformation is
\be \label{3.11}
\tilde{e}_a^j=\hat{e}_a^I\; O_I^j,\;\tilde{P}^a_j=\hat{P}^a_I\; \delta^{IJ} O_{Jj},\;
\tilde{W}^I_j=S^{IJ}\; O_{Jj},\; \tilde{V}_I^j=\delta^{mj}(V^+_{IJ}\delta^{JK} O_{Km}
+[M^{-1}]^{KL}_{mk}[V^-_{KL}-\hat{P}^a_{[K} \hat{e}_a^M\delta_{L]M}] O_{Il} \delta^{kl})
\ee
where $O=O(\theta)$. The transformation of the constraints is 
\ba \label{3.12}
&& D[u]=\int\;d^Dx\;[\hat{P}^a_I\; [L_u\hat{e}^I]_a+ 
V^+_{IJ}\; [L_u S^{IJ}] +V^-_{IJ}\; [L_u \theta^{IJ}],\;
\\
&& G[r]=\int\;d^Dx\;r^{jk} \; 
(V^+_{IJ}+V^-_{KL} [M^{-1}]^{KL}_{rs} O_I^r O_J^s) \delta^{IM}\;S^{JN}\;O_{Mj} O_{Nk}
\nonumber
\ea
The $D(D-1)/2$ Gauss constraints are naturally solved for $V^-_{IJ}$ in terms of $V^+_{IJ}, S_{IJ}$ 
and the fermionic current while the Euler angles are pure gauge degrees of freedom. 
This way the orthogonal matrix 
$O$ completely disappears from the whole description and we never need to construct it explicitly.
The $S_{IJ}$ are the searched for Gauss invariant scalars. There are $D(D+1)/2$ of them
and for the further development we need to select $D$ of them. A natural choice $X^I=1,..,D$
are e.g. the diagonal entries $X^I=S^{II}$ with conjugate momenta $Y_I=V^+_{II}$.   
A disadvantage of the polar decomposition is that solving the Gauss constraint involves the 
inverse of $S$ which can be written as $S^{-1}=\frac{\det(S) S^{-1}}{\det(S)}$ where the 
numerator is polynomial. If one aims 
at a polynomial form of the Hamiltonian constraint, then this presents an obstacle because 
$\det(S)$ is not a simple product of factors.\\
\\
{\bf Upper triangular decomposition}\\
\\
This is the advantage of the upper triangular decomposition. In this case we perform the same 
steps as in (\ref{3.9}) - (\ref{3.12}) except that $S^{IJ},V^+_{IJ}, V^-_{IJ}$ respectively are 
replaced by $U^{IJ}, R_{IJ}, L_{IJ}$ respectively where $R_{IJ}=0; I>J$ and $L_{IJ}=0; I\le J$.
In this case the Gauss constraint is solved for $L_{IJ}$ in terms of $\hat{e}_a^I,\hat{P}^a_I,
U^{IJ}, R_{IJ}$ which involves the inverse $U^{-1}$. This time we have that $\det(U)$ is a simple 
product of factors $U^{II}$ which can be exploited to write the Hamiltonian constraint as a polynomial.     
It is natural to pick $X^I:=U^{II},\; Y_I:=R_{II}$. 
\item[3.] {\bf Step 3:}\\
At the present stage we are in the following situation: Either i. the matter content provides us 
with $D$ natural scalars $X^I$ and canonical momenta $Y_I$ which do not enter the gravitational Gauss constraint and   
the gravitational phase space is still described by Gauss variant $(e_a^j, P^a_j)$. Or ii. we have performed 
steps 1., 2. above such that we are given $D$ Gauss invariant scalars $X^I$, the gravitational Gauss constraint is solved   
and the gravitational phase space is described by Gauss invariant $(\hat{e}_a^I,\hat{P}^a_I)$.
However, as far as the treatment of the SDC is concerned, both cases are identical. We will
thus consider the first case and note that all constructions that follow in steps [3.] and [4.] can also 
be applied also to $(\hat{e}_a^I,\hat{P}^a_I)$ substituted for $(e_a^j, P^a_j)$ just that 
the gravitational Gauss constraint needs to be dropped.

We will assume that the flat D-bein $h_a^I:=\partial_a X^I$ is non-degenerate. This assumption 
is of the same nature as assuming $\det(e)\not=0$ in the classical theory, except that the latter
condition is purely algebraic while $\det(h)\not=0$ depends on derivatives. Obviously such D-tuples
$X^I$ with $\det(h)\not=0$ define diffeomorphisms of $\sigma$ of some differentiabilty class,
at least unity, hence we ask how the spaces of D-tuples of arbitrary (at least once differentiable) functions     
compares in ``size'' with the space of diffeomorphisms. Again one could of course restrict to the 
cotangent space of the diffeomorphism group in the classical theory from the outset but the latter is no vector space 
but rather a very complicated Lie group not very well controlled in mathematics. To speak of ``size'' we need 
a measure on the space of functions and there appears no canonical choice. Let us instead consider the 
following subclass of degenerate D-tuples:
Suppose that $X^1,..,X^{D-1}$ are $C^1$ functions on $\sigma$ with locally linearly independent gradients.
These can be completed by some function $H$ to a (local) coordinate chart.
We ask for a function $X^D$ such that $\det(h)=0$ given $X^1,..,X^{D-1}$. By assumption we can locally 
write $X^D=G(X^1,..,X^{D-1},H)$. Then  
$\det(h)=\det(\frac{\partial(X^1,..,X^{D-1},H)}{\partial x})\;\frac{\partial G}{\partial H}=0$.
As the determinant is non-vanishing, we find the general solution $X^D=G(X^1,..,X^{D-1})$ which 
are of course rather special functions because in the local coordinates $y^1=X^1,..,y^{D-1}=X^{D-1}, y^D=H$
the function 
$X^D=G(y^1,..,y^{D-1})$ does not depend on one of the coordinates. In this case we can we can find 
an arbitrarily close (say with respect to the Schwartz space topology of smooth functions of rapid decrease in the 
smooth category) D-tuple $\tilde{X}^D=X^D+\epsilon H$ which does provide a local diffeomorphism. In thise sense 
$\det(h)=0$ is again of measure zero and we take the same attitude towards 
quantisation as for the $\det(e)=0$ case.

With these preparations out of the way we denote by $h^a_I$ the flat inverse D-Bein $h^a_I h_a^J=\delta_I^J$ and 
perform the following canonical transformation
\be \label{3.13}
\tilde{e}_I^j:=|\det(h)|^{1/2} \; h^a_I\; e_a^j,\; \tilde{P}^I_j:=|\det(h)|^{-1/2} \; h_a^I\; P^a_j,\;
\tilde{X}^I:=X^I,\; \tilde{Y}_I:=Y_I-[e_a^j\;(h^a_I P^b_j-\frac{1}{2} h^b_I P^a_j)]_{,b}
%P^a_j de_a^j=P^a_j d(|h|^{-1/2} h_a^I) e_I^j
%=P^a_j |h|^{-1/2}[dX^I,a-1/2 h_a^I\;h^c_J dX^J,c] |h|^1/2 h^b_I e_b^j
%=P^a_j [dX^I,a h^b_I e_b^j-1/2 e_a^j\;h^b_I dX^I,b] 
%=-[P^b_j h^a_I e_a^j]_{,b}+1/2[P^a_j e_a^j h^b_I]{,b} dI
\ee
It follows that $\tilde{e}_I^j,\tilde{P}^I_j$ are scalar densities of weight $\frac{1}{2}$. 
while $\tilde{X}^I, \tilde{Y}_I$ are still scalars of density $0,1$ respectively. 
It is again not difficult to see that the difference between symplectic potentials is 
exact. Despite the fact 
that (\ref{3.13}) depends on partial derivatives up to second order, we can explicitly invert 
\be \label{3.14}
e_a^j:=|\det(\tilde{h})|^{-1/2} \; \tilde{h}_a^I\; \tilde{e}_I^j,\;  P^a_j:=|\det(\tilde{h})|^{1/2} \; \tilde{h}^a_I\; \tilde{P}^a_j,\;
X^I=\tilde{X}^I,\; Y_I=\tilde{Y}_I+[(\tilde{e}_I^j h^b_J -\frac{1}{2} \tilde{e}_J^j h^b_I) \tilde{P}^J_j]_{,b}
%[e_a^j\;(h^a_I P^b_j-\frac{1}{2} h^b_I P^a_j)]_{,b} 
%[\tilde{e}_I^j h^b_J \tilde{P}^J_j-\frac{1}{2} \tilde{e}_J^j h^b_I \tilde{P}^J_j]_{,b}
\ee
where we defined $\tilde{h}_a^I=\tilde{X}^I_{,a}$. The mechanism for why this works 
is that the variable $X^I$ that is differentiated is not touched under the transformation 
and that while $e,P$ appear also differentiated in the expression for 
$\tilde{Y}$, they transform algebraically among themselves modulo factors of $h$.
To see this more transparently, suppose we have two conjugate pairs of fields in one spatial 
dimension $(X,Y),(Q,P)$ and define $\tilde{Q}:=Q/X',\; \tilde{P}:=P\;X',\; \tilde{X}:=X,\; 
\tilde{Y}:=Y-(PQ/X')'$ then this is a canonical transformation with inverse 
$Q=\tilde{Q}\;\tilde{X}',\; P=\tilde{P}/\tilde{X}',\; X=\tilde{X},\; 
Y=\tilde{Y}+(\tilde{P}\tilde{Q}/\tilde{X}')'$ as long as $X'\not=0$. In this case 
$\tilde{Q},\tilde{P}$ have density weight $-1,2$ respectively. 

If there are additional gauge boson canonical pairs  $(W_a^\alpha, V^a_\alpha)$ present 
we then supplement the canonical transformation (\ref{3.13}) by similar terms with 
$(e_a^j, P^a_j)$ replaced by $(W_a^\alpha, V^a_\alpha)$ to obtain half density valued 
$(W_I^\alpha, V^I_\alpha)$. We will not display such terms in what follows in order 
not to clutter the formulae. 

We write the constraints in the new variables (again only their affected contributions)
\ba \label{3.15}
D[u] &=&\int_\sigma\;d^Dx\; (\tilde{P}^I_j\; [L_u \tilde{e}_I^j]+\tilde{Y}_I\; [L_u \tilde{X}^I]
\nonumber\\
G[r] &=& \int_\sigma\;d^Dx\; r^{jk}\;\tilde{P}^I_{[j}\; \tilde{e}_{Ik]}
\ea
where it is understood that the Lie derivative treats both $\tilde{e},\tilde{P}$ as 
scalar densities of density weight $\frac{1}{2}$ while $\tilde{X},\tilde{Y}$ are scalar 
densities of weight $0,1$ respectively. 
\item[4.] {\bf Step 4:}\\
The last canonical transformation also equips the $X,Y$ fields with density weight $\frac{1}{2}$.
To that end we note that 
\be \label{3.16}
\det(\tilde{e})=|\det(h)|^{D/2} \;\det(h^{-1})\; \det(e)
\;\;\Rightarrow\;\; |\det(\tilde{e})|=|\det(h)|^{(D-2)/2} |\det(e)|
\ee
which means that $|\det(\tilde{e})|$ has density weight $1+(D-2)/2=D/2$. Accordingly we 
define the canonical transformation
\be \label{3.17}
\hat{e}_I^j:=\tilde{e}_I^j,\; \hat{P}^I_j:=\tilde{P}^I_j-\frac{1}{D}\tilde{Y}_J\tilde{X}^J \tilde{e}^I_j,\;
\hat{X}^I:=|\det(\tilde{e})|^{1/D} \tilde{X}^I,\;
\hat{Y}_I:=|\det(\tilde{e})|^{-1/D} \tilde{Y}_I
\ee
where $\tilde{e}^I_j \tilde{e}_I^k=\delta_j^k$. One checks exactness of the symplectic 
potential difference. This transformation has the inverse
\be \label{3.18}
\tilde{e}_I^j:=\hat{e}_I^j,\; \tilde{P}^I_j:=\hat{P}^I_j+\frac{1}{D}\hat{Y}_J\hat{X}^J \hat{e}^I_j,\;
\tilde{X}^I=|\det(\hat{e})|^{-1/D} \hat{X}^I,\;
\tilde{Y}_I=|\det(\hat{e})|^{1/D} \hat{Y}_I
\ee
Note that the transformations (\ref{3.17}) and (\ref{3.18}) would also be  
possible in a bimetric theory \cite{44} using one metric to turn the other into a 
scalar density but not both. In that respect using a metric derived from scalars is 
superior. Note also that in \cite{16}, while we used (\ref{3.13}), (\ref{3.14}) instead of
(\ref{3.17}), (\ref{3.18}) we performed a canonical transformation just in the scalar 
sector using $\hat{Y}_I=\tilde{Y}_I\;|\tilde{Y}_I|^{-1/2}, \hat{X}^I=2\tilde{X}^I|\tilde{Y}_I|^{1/2}$
with inverse $\tilde{Y}_I=|\hat{Y}_I| \hat{Y}_I, \tilde{X}^I=\frac{1}{2}|\hat{Y}_I|^{-1}$. 
The present version has the crucial advantage that the transformation is {\it linear in the momenta}.
This means that the Hamiltonian constraint keeps the important property of being at most quadratic 
in the momenta. 
  
The constraints in the new variables are
\ba \label{3.19}
D[u] &=&\int_\sigma\;d^Dx\; (\hat{P}^I_j\; [L_u \hat{e}_I^j]+\hat{Y}_I\; [L_u \hat{X}^I]
\nonumber\\
G[r] &=& \int_\sigma\;d^Dx\; r^{jk}\;\hat{P}^I_{[j}\; \hat{e}_{Ik]}
\ea
Despite the fact that it looks exactly like (\ref{3.15}) with hyphens replaced by hats,
the difference lies in the fact that applying $L_u$ to $\hat{X}^I$ now must also use the 
formula for a density of weight $\frac{1}{2}$. The Gauss constraint does not 
receive any correction because the additional term is $\propto \hat{e}_{I[j} \hat{e}^I_{k]}=\delta_{[jk]}=0$.
\item[5.] {\bf Step 5:}\\
This step is meant as an alternative to step [4.]. As long as we do not care about the polynomiality 
of the Hamiltonian constraint, we can keep step [4.] as it is. But it turns out that using [4.] one cannot 
make the HC a polynomial after multiplying by a suitable factor. This is because (\ref{3.17}) involves 
the D-th root of $\det(\tilde{e})$ and therefore the different contributions to the HC depend 
on different powers of that. The safest way to keep the option of having a polynomial HC is to make 
sure that all inverse factors that appear are products of polynomial factors. This means that one 
needs to replace $|\det(\tilde{e})|^{1/D}$ by another scalar half-density which is also Gauss invariant
and is a polynomial in the final transformed variables. This is possible using $\sqrt{|Y_I|}$ as mentioned above 
but then the polynomial version of the HC becomes a higher polynomial in the momenta. 

In case 
that we went through steps [1.], [2.] after step [3.] the phase space is now described 
by the already Gauss invariant canonical pairs $\tilde{e}_I^J, \tilde{P}^I_J$ which are both scalar half densities and say
$X^I=U^{II},\;Y_I=R_{II}$ and $U^{IJ},R_{IJ};J>I$ which are both scalar densities of weight $0,1$ respectively.
Therefore in this case we may consider the simple Gauss invariant scalar half densities $\tilde{\sigma}_I:=\tilde{e}_I^I$ to 
accomplish our goal. 
 
In case that we did not go through steps [1.], [2.] we cannot proceed like this because $\tilde{e}_I^j$ is 
not Gauss invariant. However in this case step [2.] points us into a possible direction how to proceed.
We construct the Gauss invariant scalar metric of density weight unity 
$\tilde{q}_{IJ}:=\delta_{jk}\tilde{e}_I^j\tilde{e}_J^k$ and construct a Gauss invariant, density $1/2$ valued  
upper triangular square root $u_{IJ}=0,\;I>J$ i.e. $\tilde{q}_{IJ}=\delta^{KL}\; u_{KI}\; u_{LJ}$ and the 
SO(D) matrix $O$ defined by $\tilde{e}_I^j=:O^{Jj}\;u_{JI}$. We now go through the same construction as in 
step [2.] and end up with a Gauss reduced phase space which is described by Gauss invariant 
scalar half densities $(u_{IJ},\; r^{IJ});\; I\le J$ and $(X^I, Y_I)$. The SO(D) matrix was again absorbed 
into Gauss invariant quantities and disappeared while the Gauss constraint was solved for $l^{IJ},\; I>J$ 
where $r^{IJ}+l^{IJ}=\tilde{P}^I_j O^{Jj}$. This involves the inverse of $u$ but again we have 
that $\det(u)=\prod_{I=1}^D \tilde{\sigma}_I$ is a product of polynomial factors where $\tilde{\sigma}_I:=u_{II}$. 

Therefore, to reserve the option of having a polynomial HC instead of (\ref{3.17}), (\ref{3.18}) we consider 
for example the canonical transformation employing the geometrical scalar half density 
\be \label{3.20}
\tilde{\sigma}:=\sum_I\; \tilde{e}_I^I\;\;{\sf resp.}\;\; \tilde{\sigma}:=\sum_I u_{II}
\ee
for the case that we did, resp. did not, go through steps [1.], [2.]. Then we perform the canonical 
transformation in case we performed [1.], [2.] 
\be \label{3.21}
\hat{e}_I^J:= \tilde{e}_I^J,\; \hat{P}_I^J=\tilde{P}_I^J-\tilde{\sigma}^{-1} \delta_I^J U^{KL}\;R_{KL}, \;  
\hat{U}^{IJ}:=\tilde{\sigma} \; U^{IJ},\;
\hat{R}_{IJ}:=\tilde{\sigma}^{-1} \; R_{IJ} 
\ee 
with inverse 
\be \label{3.22}      
\tilde{e}_I^J:= \hat{e}_I^J,\; \tilde{P}_I^J=\hat{P}_I^J+\hat{\sigma}^{-1} \delta_I^J \hat{U}^{KL}\;\hat{R}_{KL},\,  
\tilde{U}^{IJ}:=\hat{\sigma}^{-1} \; \hat{U}^{IJ},\;
\tilde{R}_{IJ}:=\hat{\sigma} \; \hat{R}_{IJ} 
\ee 
where $\hat{\sigma}=\sum_I \hat{e}_I^I$. Likewise without steps [1.], [2.]
\be \label{3.23}
\hat{u}_{IJ}:= u_{IJ},\; \hat{r}^{IJ}=r^{IJ}-\tilde{\sigma}^{-1} \delta^{IJ} \tilde{X}^K\;\tilde{Y}_K, \;  
\hat{X}^I:=\tilde{\sigma} \; \tilde{X}^I,\;
\hat{Y}_I:=\tilde{\sigma}^{-1} \;\tilde{Y}_I 
\ee 
with inverse 
\be \label{3.24}
u_{IJ}:= \hat{u}_{IJ},\; r^{IJ}=\hat{r}^{IJ}+\hat{\sigma}^{-1} \delta^{IJ} \hat{X}^K\;\hat{Y}_K ,\; 
\tilde{X}^I:=\hat{\sigma}^{-1} \; \hat{X}^I,\;
\tilde{Y}_I:=\hat{\sigma} \;\hat{Y}_I 
\ee 
with $\hat{\sigma}=\delta^{IJ} \hat{u}_{IJ}$. More options exist, e.g. we could instead of a single 
half density $\tilde{\sigma}$ use the individual ones $\tilde{\sigma}_I$ and define (\ref{3.21}),
(\ref{3.24}) in a component dependent way. Such options tend to increase the degree of the polynomial
version of the HC.
\end{itemize}

\subsubsection{Transformation of the Hamiltonian constraint}
\label{s3.1.3}

What remains to be done is to write the HC in the variables (\ref{3.17}) or (\ref{3.21}) or 
(\ref{3.23}). This consists in 
expressing $e,P,W,V$ in terms of $\hat{e},\hat{P},\hat{U},\hat{R}$ ([1.], [2.] used) or $e,P,X,Y$ in terms  
$\hat{e},\hat{P},\hat{X},\hat{Y}$ without Gauss reduction or $\hat{u},\hat{r},\hat{X},\hat{Y}$ 
with Gauss reduction ([1.], [2.] not used) and then substituting 
those into the HC. This can be 
done explicitly by chaining the above transformations which yields rather lengthy 
expressions for the momenta, however, these remain {\it linear} in $\hat{P},\hat{Y},\hat{R},\hat{r}$ and 
their spatial derivatives, thus the HC remains at most quadratic in the momenta.

As we intend a Fock quantisation of the theory, we ask whether it is possible to 
multiply the HC e.g. by suitable powers of $\det(e),\det(h)$ in order that it becomes
a {\it polynomial} in all $\hat{e},\hat{P},\hat{X},\hat{Y}$ etc. so that it would 
be easy to define it as a quadratic form on the Fock space by simple normal ordering. 
As we assumed that $\det(e),\det(h)\not=0$ already in the above canonical transformations, the
``rescaled'' HC is equivalent to the original one in density weight unity under that assumption. 
We will see later 
that it is also possible to work with density weight unity in Fock representations but 
the expressions become much more involved. 

It is known that the density four Hamiltonian 
constraint obtained by multiplying the density one version by $|\det(e)|^3$ is a polynomial 
in $e,p,X,Y$ but this does not mean that it is also a polynomial in $\hat{e},\hat{p},\hat{X},\hat{Y}$.
There are two potential sources of failure. The first is due to the appearance of the square root 
of $\det(h)$ in the definitioon of $\hat{e}$. If after rewriting the constraints in terms 
the transformed variables any such square root remains, then the Hamiltonian constraint 
cannot be rescaled into a polynomial. It turns out that in the density weight zero version 
each contribution to the Hamiltonian constraint just depends on rational functions (i.e. fractions of 
polynomials) which only contain even powers of $e$. Therefore the square root dependence 
$\hat{e}$ in terms of $h,e$ is not a problem. This follows from the fact that the zero weight version 
of the Hamiltonian constraint is a rational function of the spatial metric which in turn is 
a homogeneous quadratic polynomial in the $D-$Bein. The second source however is serious:
$\hat{h}$ involves the $D-$th root of $\det(\hat{e})$ and as we will see now the various 
contributions to the constraint will depend on powers of $|\det(\hat{e})|^{1/D}$ which are 
not all multiples of $D$ unless $D=2$. As $\det(\hat{e})$ is the only Gauss invariant 
polynomial that can be constructed from $\hat{e}$, this problem cannot be avoided.
In order to still define it as a quadratic form without reducing the Gauss constraint, 
one must then resort to the deformation quantisation techniques introduced in section \ref{s8}. 

To write the Hamiltonian constraint in polynomial form we thus must reduce the Gauss constraint
which opens access to Gauss invariant polynomial half densities in the transformed variables.
Then we can use the to following argument: Suppose that the Hamiltonian constraint $C$ in density zero version 
is a finite linear combination of $k$ rational functions $N_K/D_k, k=1,..,N$ 
where $N_k,D_k$ are polynomials in the transformed variables. Then $C \prod_{k=1}^N D_k$ 
is a polynomial. That polynomial version will have a very high degree in general and that 
degree maybe significantly reduced when each $D_k$ is of the form $D_k=\prod_{l=1}^M f_l^{p_{kl}}$
where $f_l,\; l=1,..,M$ are a finite number of polynomials which are relatively prime and the
$p_{kl}$ are non-negative integers. Then it will be sufficient to consider the polynomial 
$C \prod_{l=1}^M f_l^{g_l}$ where $g_l={\sf max}(\{p_{kl;k=1,..,N}\})$. In our case the role 
of the $f_l,\; l=1,2,3$ will be played by constructs made from the Gauss invariant D-Bein. \\ 
\\
In what follows we first show at which point the polynomial strategy fails without reducing
the Gauss constraint and then demonstrate how the problem is solved using the Gauss invariant  
approach. \\
\\
Case 1: Gauss constraint not reduced:\\
The first fact to note is that $\det(h)$ itself is not a polynomial in the new variables, rather 
with $\hat{h}=\partial \hat{X}$ using (\ref{3.17}) 
\ba \label{3.25}     
\det(h) &=& \det(\partial X)
=\det(\partial [\hat{X}\det(\hat{e})|^{-1/D}])
\nonumber\\
&=& \det([\partial \hat{X})-D^{-1} \hat{X} {\sf Tr}(\hat{e}^{-1}(\partial \hat{e}))]\det(\hat{e})|^{-1/D}])
=\frac{\det[\hat{h}-D^{-1} \hat{X} {\sf Tr}(\hat{e}^{-1}(\partial \hat{e}))])}{\det(\hat{e})}
\nonumber\\
&=&
\frac{\det(\hat{e})\det(\hat{h})-\det(\hat{h})\hat{h}^a_I \hat{X}^I \det(\hat{e})\hat{e}^J_j \hat{e}_{J,a}^j)}{[\det(\hat{e})]^2}   
\ea
where $\hat{e}^I_j \hat{e}_I^k=\delta_j^k$.
Since $\det(\hat{h}) \hat{h}^{-1}, \det(\hat{e}) e^{-1}$ are homogeneous polynomials in $\hat{X},\hat{e}$ 
respectively of order $D-1$ 
it follows that $\det(h)\det(\hat{e})^2$ is a homogeneous polynomial of order $2D$ in $\hat{X},\hat{e}$. Similarly
\be \label{3.26}
\det{e}=\det(|\det(h)|^{-1/2}\; h\cdot \hat{e})=|\det(h)|^{-D/2}\det(h)\det(\hat{e})  
\ee
It follows that $\det(e)|\det(h)|^{(D-2)/2}={\rm sgn}(\det(h)) \det(\hat{e})$ is a homogeneous polynomial of degree 
$D$ in $\hat{e}$ up to sign. If $\det(h)$ is nowhere vanishing then that sign is constant if the fields 
are continuous and we may restrict $X$ further assuming the sign to be positive.   

Our strategy will therefore be to write the contributions to the density zero HC, i.e. the natural density unity 
HC that appears in the Legendre transform divided by $|\det(e)|$, in terms of polynomials times powers of 
$\det(e),\det(h)$ and then to 
investigate whether we can multiply them by a common suitable power of $\det(\hat{e}),\det(h)$ in order to turn each 
of them into a polynomial 
of lowest possible degree. In order to do this we must discuss the individual contributions to the HC separately and in 
particular specify the contribution to the HC from the scalar sector. We will consider for simplicity a minimal 
Klein-Gordon coupling with vanishing potential (in particular zero rest mass). 

The simplest contribution 
to the HC is the cosmological term
$[\Lambda |\det(e)|]/|\det(e)|=\Lambda$ which is just a constant. 
%\be \label{3.27}
%|\det(e)|=|\det(h)|^{-(D-2)/2}|\det(\hat{e})|
%\ee

For the scalar derivative term we have after dividing by $|\det(e)|$
\ba \label{3.27}
&& |\det(e)|\; e^a_j e^b_k X^J_{,a} X^K_{,b}\delta_{jk}\delta_{JK}/|\det(e)|
=e^a_j e^b_k h^J_a h^K_b\delta_{jk}\delta_{JK} 
\nonumber\\
&=& |\det(h)| \hat{e}^J_j \hat{e}^K_k \delta_{jk}\delta_{JK} 
=\frac{|\det(h)|}{[\det(\hat{e})]^2} [\det(\hat{e}) \hat{e}^J_j]\; [\det(\hat{e}) \hat{e}^J_j]\;\delta_{jk}\delta_{JK}
%\nonumber\\
%&=&
%\frac{1}{|\det(e)|\; |det(h)|^{D-3}} [\det(\hat{e}) \hat{e}^J_j]\; [\det(\hat{e}) \hat{e}^J_j]\;\delta_{jk}\delta_{JK}
\ea
In view of (\ref{3.25}), in order to make (\ref{3.27}) polynomial, we would need to multiply by $[\det(\hat{e})]^{4+n}$
for some integer $n\ge 0$. This also keeps the cosmological term polynomial.
 
For the scalar kinetic term we find after dividing by $|\det(e)|$ and using (\ref{3.26}), (\ref{3.17}) 
\ba \label{3.28}
&& \frac{\delta^{IJ} \; Y_I Y_J}{|\det(e)|^2}
=|\det(h)|^{D-2}\frac{\delta^{IJ} \; [\tilde{Y}_I+..]\;[\tilde{Y}_J+..]}{|\det(\hat{e})|^2}
\nonumber\\
&=& 
|\det(h)|^{D-2}\frac{\delta^{IJ} \; [|\det(\hat{e})|^{1/D}\hat{Y}_I+..]\;[|\det(\hat{e})|^{1/D}\hat{Y}_J+..]}{|\det(\hat{e})|^2}
\ea
where the hyphens stand for additional terms that follow from the above investigations.
This is already sufficient to see that our strategy does not work out when using (\ref{3.17}) 
for $D>2$ (in particular
physical $D=3$) because (\ref{3.27}) requires the power $|\det(\hat{e})|^{2+2(D-2)-\frac{2}{D}+n}$ for some 
integer $n$ which is a fractional power and therefore destroys polynomiality of the previous terms. We conclude that using (\ref{3.17}) 
the individual terms in the HC can be written as polynomials times fractional powers of $|\det(\hat{e})|$ for any $D>2$ but 
not entirely as a polynomial.\\
\\
Case 2: Gauss constraint reduced:\\
We consider only the variant with steps [1.], [2.] not used, using steps [1.] and [2.]
leads to completely equivalent formulae with the obvious substitution of variables as indicated 
in the previous subsection. Let us thus instead use (\ref{3.23}). Then (\ref{3.25}) is replaced by  
\be \label{3.29}
\det(h)=\det(\partial [\hat{X}/\hat{\sigma}])=\det(\hat{\sigma}^{-2}[\hat{\sigma}\;(\partial \hat{X})-(\partial \hat{\sigma})\hat{X}])
=\hat{\sigma}^{-(D+1)}(\hat{\sigma}\det(\hat{h})-[\det(\hat{h})\hat{h}^a_I] \hat{X}^I \hat{\sigma}_{,a})
\ee
i.e. $\hat{\sigma}^{(D+1)} \det(h)$ is a polynomial of degree $D+1$. The crucial difference with (\ref{3.25}) is that 
in any dimension we only need an integer power of the single function 
$\hat{\sigma}$ to achieve polynomiality, moreover, the polynomial 
degree is reduced which simplifies calculations. We now reinvestigate the calculations (\ref{3.27}) and (\ref{3.28}). 

Now 
(\ref{3.27}) becomes  
\ba \label{3.30}
&& |\det(e)|\; e^a_j e^b_k X^J_{,a} X^K_{,b}\delta_{jk}\delta_{JK}/|\det(e)|
=e^a_j e^b_k h^J_a h^K_b\delta_{jk}\delta_{JK} 
\nonumber\\
&=& |\det(h)| \tilde{e}^J_j \tilde{e}^K_k \delta_{jk}\delta_{JK} 
=|\det(h)| (\hat{u}^{-1})^{IK} \;(\hat{u}^{-1})^{JL} \delta_{IJ}\delta_{KL} 
\nonumber\\
&=& \frac{|\det(h)|}{[\det(\hat{u})]^2} [\det(\hat{u}) \;(\hat{u}^{-1})^{IK}] \;[\det(\hat{u})\;(\hat{u}^{-1})^{JL}] 
\delta_{IJ}\delta_{KL} 
\ea
To make this polynomial we need to multiply by $[\det(\hat{u})]^2 \hat{\sigma}^{D+1}$ at least, thereby preserving the polynomial
character of the cosmological term. Next,  (\ref{3.28}) becomes 
\ba \label{3.32}
&& \frac{\delta^{IJ} \; Y_I Y_J}{|\det(e)|^2}
=|\det(h)|^{D-2}\frac{\delta^{IJ} \; [\tilde{Y}_I+..]\;[\tilde{Y}_J+..]}{|\det(\tilde{e})|^2}
\nonumber\\
&=& 
|\det(h)|^{D-2}\frac{\delta^{IJ} \; [\hat{\sigma}\hat{Y}_I+..]\;[\hat{\sigma}\hat{Y}_J+..}{|\det(\hat{u})|^2}
\ea
This becomes polynomial after multiplying by $\hat{\sigma}^{(D-2)(D+1)-2+n} \det(\hat{u})^{2+m} [\hat{\sigma}^{D+1}\det(h)]^2$ for some integers $n,m$, 
thereby preserving 
the polynomial character of the previous terms. To see this in full detail 
we display the full expression
\be \label{3.32a}
Y_I=
\tilde{Y}_I+[(\tilde{e}_I^j h^b_J -\frac{1}{2} \tilde{e}_J^j h^b_I) \tilde{P}^J_j]_{,b}
=\hat{\sigma}\hat{Y}_I
+[(\hat{u}_{IK} h^b_J -\frac{1}{2} \hat{u}_{JK} h^b_I)(r^{JK}+l^{JK})]_{,b}
\ee
where $l^{JK}$ is the solution of the Gauss constraint,
$l^{IJ}$ is solved via the gravitational Gauss constraint which as we observed above can 
be made polynomial after multiplying with an integer power of $\det(\hat{u})$, and 
$r^{IJ}=\hat{r}^{IJ}+\hat{\sigma}^{-1} \delta^{IJ} \hat{X}^K\;\hat{Y}_K$.
This is the basic difference: fractional and dimension dependent powers are avoided.

Next we consider the Ricci scalar term divided by $|\det(e)|$. As it can be constructed from 
$q_{ab}$ alone which is SO(D) invariant we may write it directly in terms of the upper triangular 
$D-$Bein $e_a^I=h_a^J \hat{u}_{JK} \delta^{KI}\; |\det(h)|^{-1/2}$.  
\be \label{3.33}
R[e]=2\;e^a_J e^b_K\; (\partial_{,[a}\Gamma^{JK}_{b]}[e]+\Gamma_{[aL}^J\Gamma^{LK}_{b]}[e])
\ee
where $\Gamma$ is implicitly but uniquely defined by the torsion free condition due to our use of the 
2nd order formalism
\be \label{3.34}
\partial_{[a}\;e^J_{b]}+\Gamma_{[a}^{JL}[e]\;e_{b]L}=0
\ee
This equation is homogeneous in $e$. Its solution is therefore a rational function of $e$ and its 
spatial derivatives. Explicitly one finds 
\be \label{3.35}
\Gamma_a^{JK}[e]=\frac{\gamma_{aL}^{JKbc}[e]\;(\partial_{[b} e_{c]}^L)}{\det(e)}
\ee
where $\gamma$ is a tensor of type $(2,1,1)$ antisymmetric 
in $b,c$ and $j,k$ and a 
polynomial of degree $D-1$ in $e$ without derivatives. Therefore after rewriting 
$e_a^I=h_a^J |\det(h)|^{-1/2} \hat{u}_{JK}\delta^{KI}$ we get 
\ba \label{3.36}
&& \gamma_{aL}^{JKbc}[e]=\gamma_{IL}^{JKMN}[\hat{u}]\; h_a^I\; h^b_M\; h^c_N \;|\det(h)|^{-(D-1)/2},
\nonumber\\
&& \partial_{[b} e_{c]}^L=\delta^{LQ}\; h^P_{[c}\;\partial_{b]} (|\det(h)|^{-1/2} \hat{u}_{QP})
=|\det(h)|^{-1/2} \; \delta^{LQ}\; h_{[c}^P (\partial_{b]} \hat{u}_{QP}-\frac{1}{2\;|\det(h)|} \hat{u}_{QP} 
(\partial_{b]} \det(h))) 
\nonumber
\ea
where $\gamma_{IL}^{JKMN}$ is a polynomial of degree $D-1$ in $\hat{u}$.
Since also $\det(e)=\det(\hat{u})|\det(h)|^{-D/2}$ it follows that fractional powers of $|\det(h)|$ drop out of the 
fraction, therefore $\Gamma_a^{JK}$ can be made a polynomial by multiplying it by integer powers 
of $\det(\hat{u}),\det(h), \hat{\sigma}$. For instance 
\be \label{3.37}
\det(h)\hat{\sigma}^{D+2}  \frac{[\det(h)]_{,b}}{\det(h)}=\\hat{sigma}[\hat{\sigma}^{D+1}\det(h)]_{,b}-(D+1)\hat{\sigma}_{,b}\;[\hat{\sigma}^{D+1}\det(h)],
\hat{\sigma}^D\;\det(h)\;h^a_I=\hat{\sigma} \det(\hat{h}) \hat{h}^a_I-\det(\hat{h})\hat{h}^{ab}_{IJ} \hat{X}^J \hat{\sigma}_{,b}
\ee
where $\hat{h}^I_a=\hat{X}^I_a$ and $\det(\hat{h})\hat{h}^a_I,\;\det(\hat{h})\hat{h}^{ab}_{IJ}$ are polynomials 
in $\hat{h}$ of degree $D-1,D-2$ respectively. Multiplying by twice those powers of
$\det(\hat{u}),\det(h), \hat{\sigma}$
therefore makes the curvature tensor 
polynomial. Finally $e^a_J e^b_K=h^a_M h^b_N |\det(h)| \hat{u}^{MP} \hat{u}^{MQ}\delta_{PJ}\delta_{QK}$ 
with $\hat{u}^{IK}\hat{u}_{KJ}=\delta^I_J$ can also be made polynomial by such integer powers.

Finally we consider the gravitational kinetic term which uses 
\ba \label{3.38}
&& 2\; P^{ab}=P^{(a}_j e^{b)}_k\delta^{jk}=h^a_I h^b_J|\det(h)|\;[r^{(IK}+l^{(IK}]\hat{u}^{J)L}\delta_{KL}]=:2\;h^a_I h^b_J |\det(h)|
\hat{P}^{IJ} 
\nonumber\\
&& q_{ab}=e^j_a e^k_b=|\det(h)|^{-1} \;h_a^I\; h_b^J\;\hat{u}_{IK} \hat{u}_{JL}\delta^{KL}
=:|\det(h)|^{-1} \;h_a^I\; h_b^J\;\hat{q}_{IJ}
\ea
where we think of $l^{IJ}$ solved via the gravitational Gauss constraint which as we observed above can 
be made polynomial after multiplying with an integer power of $\det(\hat{u})$ and again 
$r^{IJ}=\hat{r}^{IJ}+\hat{\sigma}^{-1} \delta^{IJ} \hat{X}^K\;\hat{Y}_K$.
The kinetic term after dividing by 
$|\det(e)|$ is given by
\be \label{3.39}
|\det(e)|^{-2}[q_{ac} q_{bd}-\frac{1}{D-1} q_{ab}\; q_{cd}] \; P^{ab} P^{cd}
=|\det(h)|^{D-2}\det(\hat{u})|^{-2}[\hat{q}_{IK} \hat{q}_{JL}-\frac{1}{D-1} \hat{q}_{IJ}\; \hat{q}_{KL}] \; \hat{P}^{IJ} \hat{P}^{KL}
\ee
which becomes polynomial after multiplying by an integer power of $\hat{\sigma},\det(h)$ and $\det(\hat{u})$.

One can investigate by similar methods that all other standard model contributions to the HC also require only integer powers of  
$\det(\hat{u}), \det(h), \hat{\sigma}$
for polynomiality and we can turn the HC into a polynomial by multiplying it by the minimal common power of all three factors.
Similar remarks hold when we use [1.] and [2.] and (\ref{3.21}). In that way reducing with respect to the gravitational Gauss 
first appears mandatory to achieve polynomiality.

\subsubsection{Reduced phase space}
\label{s.3.1.4}

In the reduced phase space approach we solve not only the Gauss constraint but also the SDC. In the presence 
of $D$ scalar fields $X^I$ this is easily accomplished by by fixing the spatial diffeomorphism gauge 
$X^I=\delta^I_a\; x^a$. In that case, we keep all fields $e_a^j, P^a_j, X^I, Y_I$ in their original 
density weight because the gauge fixing condition is perfect, there are no residual diffeomorphisms 
allowed. The SDC is solved for $Y_I=-\delta^a_I\; D_a^{NS}$ where 
$D_a^{NS}=P^b_j \;e^j_{b,a}-[P^b_j e_a^j]_{,b}+..$ are the non scalar contributions to the SDC. With the 
above methods it is 
easy to see that this can be replaced by $D_a^{NS}=r^b_I \;u^I_{b,a}-[r^b_I u_a^I]_{,b}+..$ 
modulo terms involving the gravitational Gauss constraint where $u_a^I, r^a_I=0$ when $a>I$ with 
$e_a^j=u_a^I O_I^j$ and $O\in$ SO(D). This version of the upper 
triangular gauge condition breaks of course diffeomorphism covariance as it mixes tensor and 
scalar label indices but this is precisely correct as we are fixing the diffeomorphism gauge anyway. 
We may then substitute this solution for $Y_I$ and the solution of the gravitational Gauss
constraint for $l_a^I$ with  $l_a^I=0$ when $a\le I$, into the HC. A polynomial version can now be 
found much simpler as $h_a^I=\delta_a^I$ is a constant and background independent 
Fock representations such as based on the annihilator $u_J^I-i r^J_I$ are available where 
$u_J^I=\delta^a_J u_a^I, r^J_I=r^a_I \delta_a^J$. Note that this is truly background independent 
because we have identified scalar and tensor labels.

To see this more explicitly we can also construct the relational Dirac observables with respect 
to the SDC. These are simply the pull-backs of all fields by the ``dynamical'' diffeomorphism
$x\mapsto y=X(x)$ such as 
\be \label{3.40}
[X^\ast e]_I^j=h^a_I\; e_a^j\circ X^{-1},\; 
[X^\ast P]^I_j=[\frac{h^I_a\; P^a_j}{\det(h)}]\circ X^{-1} 
\ee
These have vanishing Poisson brackets with the SDC because we do not only map $e,P$ to scalars 
of density weight zero with respect to $x$ using $h_a^I=X^I_{,a}$ but moreover we also transform 
the argument using $X$. Additionally, they have canonical brackets on the target space of the 
dynamical diffeomorphism and the contribution to the gravitational Gauss constraint is simply 
$[X^\ast P]^I_{[j} [X^\ast e]_{Ik]}$. Obviously this Dirac observable and gauge fixing viewpoint are 
completely equivalent and the gauge fixed variables may therefore be viewed as target space 
scalar densities of weight $1/2$ because there are no diffeomorphisms allowed any longer on the 
target space.     

The HC gauge freedom may also be gauge fixed 
and an explicit, non-perturbative physical Hamiltonian results if there are additional scalar fields, see e.g. \cite{21}.
In the present case we can choose them as one of the vector bosons not gauge fixed yet or the 
Higgs field itself. This is possible because the momenta of these scalar fields do enter the constraints
without spatial derivatives.

\subsection{Background independent Fock representations}
\label{s3.2} 

The available background independent Fock representations based on the 
scalar density weight $1/2$ conjugate scalars  derived in the previous 
subsection depend on whether we use [1.], [2.]
of the previous subsection or not and whether we care about the existence of polynomial
versions of the HC.\\
\\
We consider the most complicated case, i.e. that [1.], [2.] are not used and that polynomiality of the HC is not an issue. 
We concentrate on the geometry sector, the considerations for the scalar sector are analogous but simpler because 
it does not contribute to the Gauss constraint. The considerations for the case that [1.], [2.] 
are used are also simpler because there is no longer any Gauss constraint.
We thus construct the annihilators for the geometry sector
\be \label{3.41}
a_I^j=\frac{1}{\sqrt{2}}[\kappa_{IK}\delta^{KJ}\;\hat{e}_J^j-i\; \delta_{IK}\;(\kappa^{-1})^{KL}\; \delta_{LJ} \hat{P}^J_k \delta^{jk}]
\ee       
which obey the canonical commutation relations (CCR)
\be \label{3.42}
[a_J^j(x),a^K_k(x')]=\delta_{jk}\delta^{JK}\delta(x,x')
\ee
The positive, thus symmetric, invertible, constant ``Immirzi'' matrix $\kappa_{IJ}$ presents
a $D(D+1)/2$ parameter degree of freedom. It cannot be made position dependent or dependent on 
the Gauss indices without breaking covariance. In what follows we consider the simplest 
choice $\kappa_{IJ}=\delta_{IJ}$ for simplicity. Likewise
\be \label{3.43}
a^I=\frac{1}{\sqrt{2}}[\hat{X}^I-i\delta^{IJ}\hat{Y}_J]
\ee
modulo a similar possible $\kappa$ freedom. One might also consider more general annihilators 
based on complex linear combinations of all $\hat{e},\hat{P},\hat{X},\hat{Y}$ but this would break 
Gauss covariance.     

If [1.], [2.] are not used but polynomiality of the HC is an issue then instead of (\ref{3.41}) we 
consider 
\be \label{3.44}
a_{IJ}=\frac{1}{\sqrt{2}}[\kappa_{IK}\delta^{KJ}\;\hat{u}_{IJ}-i\; \delta_{IK}\;(\kappa^{-1})^{KL}\; \delta_{LM} \hat{r}^{MN} \delta^{NJ}]
\ee        
with $\hat{u},\hat{r}$ upper triangular
while (\ref{3.42}) is kept as it is. In this case we can consider further possibilities of mixing
$\hat{u}, \hat{r},\hat{X},\hat{Y}$ because the Gauss constraint is already solved classically. In general the 
freedom in choosing Fock representations increases with the amount of constraints that are already solved 
classically. 

Finally, if [1.], [2.] are used then the Gauss constraint is automatically soved and we use the 
annihilator 
\be \label{3.45}
a_{IJ}=\frac{1}{\sqrt{2}}[\kappa_{IK}\delta^{KJ}\;\hat{e}_{IJ}-i\; \delta_{IK}\;(\kappa^{-1})^{KL}\; \delta_{LM} \hat{P}^{MN} \delta^{NJ}]
\ee   
while (\ref{3.42}) is kept. The difference with (\ref{3.44}) is that here we solve the Gauss constraint classically using
part of the vector boson sector rather than the geometry so that $\hat{e}, \hat{P}$ are general rather upper triangular.

\section{Gauss and spatial diffeomorphism constraint operators}
\label{s4}
 
We consider for the sake of generality the case that the Gauss constraint is still present,
otherwise apply the present section only to the SDC. \\
\\
Recall that we have the constraints (we display only the bosonic contributions for simplicity)
\be \label{4.1}
G[r]=\int\; d^Dx\; r^{jk} \hat{P}^I_j \; \delta_{kl}\hat{e}_I^l,\\;
D[u]=\frac{1}{2}\int\; d^Dx\; u^a\; [\hat{P}^I_j\hat{e}_{I,a}^j-\hat{P}^I_{j,a} \hat{e}_I^j]
\ee
Using (\ref{3.41} with $\kappa_{IJ}=\delta_{IJ}$ we can rewrite these in terms of annihilators and 
creators
\be \label{4.2}
a_J^j=\frac{1}{\sqrt{2}}[\hat{e}_J^j-i\delta_{JK}\delta^{jk} \hat{P}^K_k],\;\;
\hat{e}_J^j=\frac{1}{\sqrt{2}}[a_J^j+(a_J^j)^\ast],\;\;
\hat{P}^J_j=\frac{i}{\sqrt{2}}\delta^{JK}\delta_{jk}[a_K^k-(a_K^k)^\ast],\;\;
\ee
This leaves an ordering ambiguity. We choose normal ordering
\ba \label{4.3}   
G[r] &=& -i\int\; d^Dx\; r_{jk} \;\delta^{JK}\; [a_J^j]^\ast \; a_K^k,\;\;
\nonumber\\
%P\times e=i/2 (a-a*)\times(a+a*)=i/2(-2 a*\times a)=-ia*\times a 
D[u] &=& -\frac{i}{2}\int\; d^Dx\; u^a\; \delta_{jk}\delta^{JK}\;([a_J^j]^\ast\; a_{K,a}^k-
[a_{J,a}^j]^\ast\; a_K^k)
%i/4[(a-a*)(a+a*)'-(a-a*)'(a+a*)]=i/4[aa'+a a*'-a* a'-a* a*'-a'a - a' a* + a*' a + a*' a* =i/2(a*' a-a* a')
\ea
The crucial feature of these expressions is that there are no terms quadratic in the creators as 
such terms would map each vector in the span of Fock states outside the Hilbert space due to 
both an IR and UV divergence. This was 
the whole purpose of the previous section: In no other Fock algebra would that be the case, for 
any alternative choice with non-constant $\kappa$ there would not have been the corresponding 
cancellations when expanding $2^{1/2}\hat{e}=a+a^\ast,\;
2^{1/2}\hat{P}=i(a-a^\ast)$. In reordering (\ref{4.2}) we used $[a,a]=[a,\partial a]=0$ and 
adjoints of those.   

Now notice that when $[a_\alpha(x),(a_\beta(y))^\ast]=\delta_{\alpha\beta}\delta(x,y)$, all others vanishing, 
then the normal ordered bi-linear quadratic forms  $B_{\alpha\beta}(x):=[a_\alpha(x)]^\ast\; a_\beta(x)$ 
obey the formal closed commutator algebra
\be \label{4.4}
[B_{\alpha\beta}(x),B_{\gamma\delta}(y)]=\delta(x,y)
[\delta_{\beta\gamma}\; B_{\alpha\delta}(x) 
-\delta_{\delta\alpha}\; B_{\gamma\beta}(x)] 
\ee
preserving normal order. This already implies that the formal commutator algebra of the 
expressions (\ref{4.2}) delivers automatically normal ordered expressions without producing
normal ordering constants that would lead to anomalies. 

Even more is true: we show that $G[r], D[u]$ are densely defined, symmetric
operators on the domain ${\cal D}_0$ of 
Fock states $a[f_1]^\ast..a[f_N]^\ast\Omega$ with Fock vacuum $a_J^j(x)\Omega=0$ and 
$f_1,..,f_N$ smooth, complex valued smearing functions, say of rapid decrease and where 
\be \label{4.4a}
a[f]=\int\; d^Dx\; [f^I_j]^\ast\;a_I^j     
\ee
In addition, these can be uniquely extended to self-adjoint operators. To see this 
we start by observing $G[r]\Omega=D[u]\Omega=0$ and the easy calculations 
based on the elementary CCR
\be \label{4.5}
[a_J^j(x),(a_K^k)^\ast(y)]=\delta_{JK}\delta^{jk}\delta(x,y)
\ee
given by
\ba \label{4.6}
[G[r],a[f]^\ast]
%=-i\int\; d^Dx\int d^Dy\;r_{jk}(x)\; f^L_l(y)\;\delta^{JK}\;[a_J^j(x)]^\ast\;[a_K^k(x),[a_L^l]^\ast(y)]
&=&-i\int\; d^Dx\; r_{jk}\;\delta^{kl}\; f^J_l\; a_J^k=-i\;a[r\cdot f]^\ast
\nonumber\\
{[}D[u],a[f]^\ast]
%=-\frac{i}{2}\int\; d^Dx\int d^Dy\;u^b(x)\; f^L_l(y)\;\delta^{JK}\;\delta_{jk}\;
%([a_J^j(x)]^\ast\;[a_{K,b}^k(x),[a_L^l]^\ast(y)]
%-([a_{J,b}^j(x)]^\ast\;[a_K^k(x),[a_L^l]^\ast(y)])
%=-\frac{i}{2}\int\; d^Dx\int d^Dy\;u^b(x)\; f^L_l(y)\;\delta^{JK}\;\delta_{jk}\;\delta_{KL}\delta^{kl}\;
%([a_J^j(x)]^\ast\delta_{,x^b}(x,y)
%-([a_{J,b}^j(x)]^\ast\delta(x,y))
%=\frac{i}{2}\int\; d^Dx\;f^J_j \{(u^b [a_J^j]^\ast)_{,b}+u^b [a_{,b}J^j]^\ast\}
&=& -\frac{i}{2}\int\; d^Dx\;\{(f^J_{j,b}\;u^b+(f^J_j\;u^b)_{,b}\} 
=-i\;a[L_u f]^\ast
\ea
where 
\be \label{4.7}
[r\cdot f]^J_j=r_{jk}\delta^{kl}\; f^J_j,\;
[L_u\; f]:=u^b\; f^J_{j,b}+\frac{1}{2}\; u^b_{,b} \; f^J_j
\ee
are the natural actions of the Lie algebras so(D) and diff$(\sigma)$ (the vector fields on $\sigma$)
of the gauge groups SO(D) and Diff$(\sigma)$ respectively on the space of functions $f$ 
which are valued in the D-dimensional vector representation of SO(D) and the density weight $\frac{1}{2}$
scalars. In order to obtain (\ref{4.6}) we had to integrate by parts which is justified relying 
on the rapid decrease of the functions $f$. 

We lift the action of the Lie algebra of the gauge groups on the Fock space to 
the full Lie group ${\sf Diff}(\sigma) \ltimes  {\cal G}$ which consists of the 
semi-direct product of local SO(D) rotations and spatial diffeomorphisms by defining 
for SO(D) valued functions $g$ on $\sigma$ and spatial diffeomorphisms $\varphi$ the 
operators $U(g),U(\varphi)$ via 
\be \label{4.8}
U(g)\; a[f_1]^\ast\;..\;a[f_N]^\ast\;\Omega:=a[g\cdot f_1]^\ast\;..\;a[g\cdot f_N]^\ast\Omega,\;
U(\varphi)\; a[f_1]^\ast\;..\;a[f_N]^\ast\;\Omega:=a[\varphi\cdot f_1]^\ast\;..\;a[\varphi\cdot f_N]^\ast\Omega
\ee
where 
\be \label{4.9}
[g\cdot f]^J_j(x)=\delta_{jl}\; [g^{-1}]^{lk}(x)\; f^J_k(x),\;
[\varphi\cdot f]^J_j(x)=|\det([\frac{\partial\varphi^{-1}}{\partial x}])|^{1/2}\; f^J_k(\varphi^{-1}(x))
\ee
which is a local rotation in the D-dimensional vector representation and the pull-back by $\varphi^{-1}$ 
of a scalar half density. 

It follows that
%This interpretation of the test functions $f$ also explains why 
the inner product between Fock 
states 
\be \label{4.10}
<a[f_1]^\ast..a[f_M]^\ast\Omega,\;a[g_1]^\ast..a[g_N]^\ast\Omega>_{{\sf Fock}}
=\delta_{MN}\; \sum_{\pi \in S_N}\; \prod_{A=1}^N\; <f_A,g_{\pi(A)}>_1
\ee
is both SO(D) and Diff$(\sigma)$ invariant under (\ref{4.9}). Here 
\be \label{4.11}
<f,g>_1=\delta_{JK}\delta^{jk}\int_\sigma \; d^Dx\; [f^J_j]^\ast \; g^K_k
\ee
is the 1-particle inner product on the space of test functions which 
completes to the Hilbert space $\mathfrak{h}_1$ and which is manifestly 
invariant under both gauge groups precisely because $f,g$ are half densities. 
If they were not, one would need a background volume form to define the 1-particle 
inner product which would break background independence and spatial diffeomorphism invariance.

It follows that the operators (\ref{4.8}) are isometric on the dense domain ${\cal D}_0$ 
and have the inverse $U(g)^{-1}=U(g^{-1}),\; U(\varphi)^{-1}=U(\varphi^{-1})$ there. They 
therefore extend to the full Fock space as unitary operators by continuity. 
Consider the one parameter subgroups generated 
by so(D) valued functions $r$ and vector fields $u$  
\be \label{4.10a}
g^r_t(x)=\exp(t r(x)),\; r(x)=r_{jk} E^{jk},\; [E^{jk}]_{mn}=\delta^j_{[m}\delta^k_{n]};\;\;
\varphi^u_t(x)=c^u_x(t)
\ee
where $c^u_x(t)$ is the integral curve of $u$ through $x$, that is in local coordinates, 
$\dot{c}^u_x(t)=u(c^u_x(t)),\;c^u_x(0)=x$. Note the group property $g^r_s \;g^r_t=g^r_{s+t},\;
\varphi^u_s\circ \varphi^u_t=\varphi^u_{s+t}$.
Then  
\be \label{4.11a}
\{\frac{d}{dt} [g^r_t\cdot f]^J_j\}_{t=0}=-[r\cdot f]^J_j,\;
\{\frac{d}{dt} [\varphi^u_t\cdot f]^J_j\}_{t=0}=-[L_u\; f]^J_j
\ee
Next, the one parameter unitary groups $t\mapsto U(g^r_t),\; U(\varphi^u_t)$ are strongly continuous 
\be \label{4.12}
\lim_{t\to 0}||[U(g^r_t)-1_{{\cal H}_{{\sf Fock}}}]\; a[f_1]^\ast..a[f_M]^\ast\Omega||=0
\ee
and similar for $U(\varphi^u_t)$. They therefore have self-adjoint generators 
$A[r], B[u]$ respectively defined by 
\be \label{4.13}
U(g^r_t)=e^{it\;A[r]},\; U(\varphi^u_t)=e^{it\;B[u]} 
\ee 
via Stone's theorem \cite{46}. 

The domain of self-adjointness ${\cal D}_r,\; {\cal D}_u$ of $A[r],\; B[u]$ 
respectively consists of those $\psi\in {\cal H}_{{\sf Fock}}$ 
for which the strong derivative of $U(g^r_t)\psi,\;U(\varphi^u_t)\psi$ at $t=0$ exists. 
These domains include ${\cal D}_0$. For instance 
\ba \label{4.14}
&& i\;A[r]\; a[f]^\ast \; \Omega=\{\frac{d}{dt}\; U(g^r_t) \; a[f]^\ast \; \Omega\}_{t=0}
=\{\frac{d}{dt}\; a[g^r_t\cdot f]^\ast \; \Omega\}_{t=0}=-a[r\cdot f]^\ast \Omega,\;
\nonumber\\
&& i\;B[u]\; a[f]^\ast \; \Omega=\{\frac{d}{dt}\; U(\varphi^u_t) \; a[f]^\ast \; \Omega\}_{t=0}
=\{\frac{d}{dt}\; a[\varphi^u_t\cdot f]^\ast \; \Omega\}_{t=0}=-a[L_u\; f]^\ast \Omega
\ea
Comparing (\ref{4.14}) and ({\ref{4.6}) we see that $A[r],\; B[u]$ are the 
unique self-adjoint extensions of $G[r], \; D[u]$ obtained from the strongly 
continuous 1-parameter groups. We will not distinguish between 
$G[r],\; A[r]$ and $D[u],\;B[u]$ in what follows.

Likewise
we can compute the corresponding Fock vacuum expectation value state on the Weyl elements 
\be \label{4.15}
W(f,g)=\exp(i[<f,\hat{e}>_1+<g,\hat{P}>_1])
\ee
for real valued $f,g$ (we are using Kroneckers to move index position) by expressing
the Weyl element in terms of $a,a^\ast$, reordering it into normal order and using 
the vacuum property   
\be \label{4.16}
\omega(W(f,g))=\exp(-\frac{1}{4}[<f,f>_1+<g,g>_1])
\ee
We may therefore think of ${\cal H}_{{\sf Fock}}$ both as GNS Hilbert space determined 
by this state with cyclic vacuum $\Omega$ or as the space of square integrable 
functions $L_2(T,d\mu)$ where $T$ is the space of tempered distributions and $\mu$ 
the Gaussian measure on $T$ with white noise covariance $\kappa={\sf id}_{\mathfrak{h}_1}$
\cite{45}. Note that the state (\ref{4.11}) is manifestly invariant under both 
gauge groups. 

\section{The hypersurface deformation algebroid}
\label{s5}

We note that both $G[r], \; D[u]$ preserve the common domain ${\cal D}_0$. We may therefore 
compute their algebra of commutators. For instance 
\ba \label{5.1}
&& [G[r],G[s]]\; a[f]^\ast\;\Omega=-a[(r\cdot s-s\cdot r)\cdot f]^\ast\Omega=
i G[-[r,s]]\; a[f]^\ast\Omega
\nonumber\\
&& [D[u],G[s]]\; a[f]^\ast\;\Omega
=-a[ L_u \;(r\cdot f)-r\cdot (L_u\; f)]^\ast\Omega
=-a[ (L_u \;r)\cdot f]^\ast\Omega
=i G[-(L_u \; r)]\; a[f]^\ast\Omega   
\nonumber\\
&& [D[u],D[v]]\; a[f]^\ast\;\Omega
=-a[(L_u \;L_v-L_v\; L_u)\;f]^\ast\Omega
=-a[ (L_{[u,v]}\;f]^\ast\Omega
=i D[-[u,v]]\; a[f]^\ast\Omega   
\ea
where $[r,s]$ is the commutator of antisymmetric matrices, $L_u \;r$ the Lie derivative 
of the density weight zero scalar matrix $r$ and $[u,v]$ the commutator of vector fields 
considered as derivations on the space of smooth functions. Comparing with 
the classical Poisson bracket relations
\be \label{5.2}
\{G[r],G[s]\}=G[-[r,s]],\;\{D[u],G[s]\}=G[-L_u\; r],\; \{D[u],D[v]\}=D[-[u,v]]
\ee
we see that the Gauss and SDC are implemented {\it without anomalies} in the 
quantum theory in the sense that their commutators reproduce the result of the classical Poisson 
bracket times $i$. 
 
This result can be traced back to our background independent starting point. Recall that 
in 1+1 dimensions, say parametrised field theory on the cylinder \cite{47}, one 
can construct from SDC $D$ and HC $C$ two linear combinations $V_\pm=D\pm C$ whose Poisson 
brackets generate the direct sum of Lie algebras diff$(S^1)\oplus$diff$(S^1)$. The 
usual Fock quantisations of those however use a background metric on $S^1$ as these are 
defined using Fourier modes and mix scalars of different density weight. 
One finds that $V_\pm$ close up to central charges $c_\pm$. 
Technically these arise because when expanding $V_\pm$ in terms of annihilation and 
creation operators, there are terms of the form $a^2,(a^\ast)^2$ and not only $a^\ast a$ as 
in our case. The commutator calculation then includes terms 
$[a^2,(a^\ast)^2]\propto (a a^\ast+a^\ast a)=2 a^\ast a+c$. Hence, the absence of a central 
charge in our case has a simple explanation and is again due to the absence of those $(a^\ast)^2$
terms.

Finally we include the HC into the analysis. The details are left for a future 
publication, however, the analysis of the previous section shows that we will be able 
to quantise it in its smeared form $C[N]$ as a normal ordered quadratic form densely defined 
either on ${\cal D}_0$ when $C$ is polynomial or on some other dense domain based on 
excitations of coherent states in the non-polynomial case as we will see in later sections. 
In both cases the integrand of $C$ is a manifestly 
rotation invariant scalar density of some density weight $w$ whose terms are 
polynomials in annihilation and creation operators in the polynomial case and 
in addition of their exponentials in the non-polynomial case. As $U(g), U(\varphi)$ preserve 
the domain of the quadratic form, the objects $U(g) \; C[N]\;U(g)^{-1}, \;  
U(\varphi) \; C[N]\;U(\varphi)^{-1}$ are again well defined quadratic forms on the same 
domain. But by construction $U(g)\; C[N]\; U(g)^{-1}=C[N]$ and $U(\varphi)\; C[f]\; U(\varphi)^{-1}=
C[\varphi\cdot N]$ where $\varphi\cdot N$ is the action of $\varphi^{-1}$ on the space of scalar 
densities of weight $1-w$ valued lapse functions. We can now consider the above 1-parameter unitary groups 
and conclude 
\be \label{5.3}
[G[r],C[N]]=0,\;\;[D[u],C[N]]=i\;C[-L_u\; N]
\ee
thereby reproducing the classical Poisson brackets $\{G[r],C[N]\}=0,\; \{D[u],C[N]\}=-C[L_u\; N]$.
In more detail consider a polynomial term of the form
\be \label{5.4}
\int\; d^Dx\; N\; I_{j_1...j_{M+N}}\;[\prod_{A=1}^M \;(a_{J_A}^{j_A})^\ast]\; [\prod_{B=M+1}^{M+N} \;a_{J_B}^{j_B}]   
\ee
where $I$ is a gauge invariant intertwiner between the $M+N$ fold tensor product of defining representations 
of SO(D) and the trivial representation. Since 
\be \label{5.5}
U\; [a^\ast]^M\; a^N\; U^{-1}=[(U\;a\; U^{-1})^\ast]^M\; [U\; a\; U^{-1}]^N,\;
U(g)\;a\; U(g)^{-1}=g\cdot a,\; U(\varphi)\;a\; U(\varphi)^{-1}=\varphi\cdot a
\ee
where the latter is the action of $\varphi$ on the scalar half density $a$, the claim (\ref{5.3})
immediately follows. The same is true if $C$ includes exponentials of $a, a^\ast$ as 
$U \exp(z\;a) U^{-1}=\exp(z\;U\;a\; U^{-1})$ for any complex coefficient $z$.\\ 
\\
We conclude that in the present half-density Fock representation, the only possible 
source of anomalies 
in the HDA resides in the commutator $[C[N],C[N']]$ for two different reasons: First, because 
one needs to give meaning to that calculation as quadratic forms cannot be multiplied. Second,
because $C$ is much more complicated than $G,D$ and not only contains $a^\ast\;a$ terms but 
also terms that after performing the commutator are not automatically normal ordered. Bringing those into normal
order potentially produces anomalies which are no longer central because $C$ is not a quadratic 
polynomial. See \cite{17} for an illustration of both effects in a much simpler, polynomial but closely related 
setting. In \cite{17} it is shown that one can give meaning to the commutator as a weak limit 
of regulated Hamiltonian constraints and the regulator can be chosen such that no anomalies arise. 
The corresponding analysis in the present case is very complex and must be deferred to a different 
work.

\section{Generalised zero eigenvectors}
\label{s6}

We define a generalised zero eigenvector as a linear functional $l$ on the span ${\cal D}_0$ of Fock states 
which solves all constraints in the sense that
\be \label{6.1}
l[G[r]\psi]=l[D[u]\psi]=l[C[N]\psi]=0
\ee
for all $\psi\in {\cal D}_0$ and all $r,u,N$. Note that these conditions for $G,D$ are automatically 
well defined as $l:\; {\cal D}_0\to \mathbb{C}$ and $G[r]\;{\cal D}_0,\; D[u]\;{\cal D}_0 \subset {\cal D}_0$ 
as we saw in the previous section. However $C[N]$ is just a quadratic form and $C[N]\psi$ is not defined as an
element of the Fock space, we can just compute matrix elements $<\psi, C[N]\psi'>,\; \psi,\psi'\in {\cal D}_0$.

However, we note that ${\cal D}_0$ contains the elements of the orthonormal basis 
\be \label{6.2}
b_{\{n\}}:=\prod_\alpha\; \frac{[a_\alpha^\ast]^{n_\alpha}}{\sqrt{n_\alpha !}}\;\Omega
\ee
where $n_\alpha\in \mathbb{N}_0$ equals zero for all but finitely many of the $\alpha$ which 
run through the countable mode label set $\cal A$ of an orthonormal basis $e_\alpha$ of the 
separable 1-particle Hilbert space which define $a_\alpha:=a[e_\alpha]$. The $e_\alpha$ can 
be chosen smooth of rapid decrease.
Then $l$ is uniquely 
specified by the numbers $l_{\{n\}}:=l[b_{\{n\}}]$ and we have the identity
\be \label{6.3}
l[.]=\sum_{\{n\}}\; l_{\{n\}}\; <b_{\{n\}}, .>_{{\cal H}}
\ee
by linearity. Accordingly 
\be \label{6.3a}
l[C[N]\psi]=\sum_{\{n\}}\; l_{\{n\}}\; <b_{\{n\}}, C[N]\psi>
\ee
is computable and requiring it to vanish for all $N,\psi$ is an infinite system of conditions 
for the coefficients $l_{\{n\}}$. In fact it is equivalent to require that 
\be \label{6.4}
l[C[N]\;b_{\{n\}}]=\sum_{\{m\}}\; l_{\{m\}}\; <b_{\{m\}}, C[N]\;b_{\{n\}}>
\ee
for all $\{n\}=\{n_\alpha\}_{\alpha\in {\cal A}}$ and all $N$.

In what follows we study exact solutions to $G[r]=D[u]=0$. Exact solutions that will additionally 
solve $C[N]=0$ are harder to find and will be subject of other publications. Consider 
the infinite set of quadratic forms 
\be \label{6.5}
Q_{J_1 K_1..J_N K_N}:=
\prod_{A=1}^N \;Q_{J_A K_A},\;\; Q_{JK}:=\int\; d^Dx\; \delta_{jk}\; [a_J^j]^\ast \; [a_K^k]^\ast
\ee
We claim that the not normalisable vector 
\be \label{6.6}
l_{J_1 K_1..J_N K_N}:=Q_{J_1 K_1..J_N K_N}\;\Omega
\ee
lies in the joint generalised kernel of both $G[r], D[u]$ for all $r,u$. The action of 
(\ref{6.6}) on ${\cal D}_0$ is by taking the formal adjoint of (\ref{6.5})
\be \label{6.7}
l_{J_1 K_1..J_N K_N}[\psi]=<\Omega,\prod_{A=1}^N Q_{J_A K_A}^\ast \psi>,\;\;
Q_{JK}^\ast:=\int\; d^Dx\; \delta_{jk}\; a_J^j \; a_K^k     
\ee
We have 
\ba \label{6.8}
&& [G[r],\; Q_{JK}^\ast]=-i\int\; d^Dx\; \delta_{jk}\;\{(r\cdot a_J)^j\; a_K^k+a_J^j\;(r\cdot a_K)^k\}
=-i\int\; d^Dx\;\delta_{jk}\;r_{lm} \{\delta^{jl}\; a_J^m\; a_K^k+\delta^{kl}\; a_J^j\; a_K^m\}
\nonumber\\
&=& -i\int\; d^Dx\;r_{lm}\{a_J^m\; a_K^l+a_J^l\; a_K^m\}   
=0
\\
&& [D[u],\; Q_{JK}^\ast]=-\frac{i}{2}\int\; d^Dx\; \delta_{jk}\;\{(L_u \; a_J^j)\; a_K^k+a_J^j\;(L_u a_K^k)\}
=-\frac{i}{2}\int\; d^Dx\; \delta_{jk}\;\{(L_u \; a_J^j)\; a_K^k+a_J^j\;(L_u\; a_K^k)\}  
\nonumber\\
&=&-\frac{i}{2}\int\; d^Dx\; \delta_{jk}\;\{L_u\; (a_J^j\; a_K^k)\}
=-\frac{i}{2}\int\; d^Dx\; \delta_{jk}\;\{u^b\; (a_J^j\; a_K^k)\}_{,b}
\nonumber
\ea
The first commutator is manifestly vanishing as we contract an object antisymmetric in $j,k$ 
with a symmetric object. The second commutator acting on a Fock vector with particle number at 
least two is not obviously vanishing but applying it to such a Fock vector we obtain contributions 
proportional to
\be \label{6.9}
\int\; d^Dx\; \delta^{jk}\;\delta_{JM}\delta_{KN}\; \{u^b f^M_j\; \tilde{f}^N_k)\}_{,b}=0
\ee
as $f,\tilde{f}$ are of rapid decrease. Accordingly we can pull $G[r],D[u]$ through the $Q_{J_A K_A}$ 
and then use\\
 $<\Omega,G[r]\psi>=<\Omega,D[u]\psi>=0$ for $\psi\in {\cal D}_0$ as $G[r],D[u]$ 
contains only $a^\ast a$ terms and 
$\prod_A Q_{J_a K_A}^\ast$ preserves ${\cal D}_0$. The latter statement follows from the 
fact that 
\be \label{6.10}
Q_{JK}^\ast a[f_1]^\ast..a[f_N]^\ast\Omega=\sum_{A\not=B}\; c^{AB}_{JK}\; \prod_{C\not=A,B}\; a[f_C]^\ast \Omega,\;
c^{AB}_{JK}=\int d^Dx\; \delta^{jk}\; \delta_{JM}\delta_{KN}\; [f_A]^M_j\; [f_B]^N_k
\ee
and the coefficients involved obey 
\be \label{6.11}
|c^{AB}_{JK}|\le \delta^{jk}\; ||[f_A]^J_j||_{L_2(\sigma)}\;[f_B]^K_k||_{L_2(\sigma)}
\le D\; ||f_A||_1\; ||f_B||_1
\ee
due to the Cauchy-Schwarz inequality.\\  
\\
Thus we obtain an infinite number of non trivial solutions $l$ in the joint generalised kernel
of $G[r]=D[u]=0$. It is easy to see that except for $l=<\Omega,.>$ none of these solutions are 
normalisable. If we compare to the LQG incarnation, there are two key differences: First, in 
the LQG incarnation solutions of just the Gauss constraint are still normalisable. Here 
a class of solutions to just the Gauss constraint can be constructed as polynomials of the 
\be \label{6.12}  
Q[F]=I_{j_1..j_N}\;\int\; d^Dx \; F^{J_1..J_N}(x)\;\prod_{A=1}^N\; [a_{J_A}^{j_A}(x)]^\ast 
\ee
applied to $\Omega$ where as in the previous section $I$ is a gauge invariant 
intertwiner and $F$ an arbitrary smooth function of rapid decrease. 
It is easy to see that for $N>1$ such vectors are not normalisable and for 
$N=1$ they cannot be rotation invariant. Thus such solutions to just 
the Gauss constraint in the present Fock incarnation are not normalisable. The second 
difference is that joint solutions of $G=D=0$ in the LQG incarnation required an
uncountable  
orbit sum over graphs. Here such uncountable sums are avoided, the solutions can be concretely 
constructed using familiar Fock techniques.

Note that $l_{J_1..K_N}$ is a linear combination of the
\be\label{6.13}
:\;\prod_A Q'_{J_A K_A}\;:\;\Omega, \;\;Q'_{JK}=\int\; d^Dx\; \hat{e}^j_J \hat{e}^k_K
\ee
where the $Q'_{JK}$ are the spatial average over the scalar components of the metric which 
offers a partial interpretation of these solutions. 
Certainly the solutions $l_{I_1..J_N}$ are not the most general ones. The common idea 
is to construct them from integrals of densities of weight unity. This 
forbids integrals over polynomials of degree other than two if no 
derivatives are involved. However, assuming that we can give meaning to inverse fractional powers of 
$\det(\hat{e})$ we may attempt to generate solutions from normal ordered 
polynomials of the 
\be \label{6.14}
Q_{J_1..J_N}=\int\; d^Dx\; I_{j_1.. j_N}\;\frac{\prod_{A=1}^N\; \hat{e}_{J_A}}{|\det(\hat{e})|^{(N-2)/D}}
\ee
The quest for giving meaning to $|\det(\hat{e})|^t,\; t\in \mathbb{R}$ leads us to the next two sections.

\section{Domains, coherent and squeezed states}
\label{s7}

An elementary property of Fock representations is that they are irreducible for the underlying 
Weyl algebra \cite{28}. Applied to our concrete setting (which can be genneralised), 
given a 1-particle Hilbert space $\mathfrak{h}$ 
and the Weyl elements $W(f,g),\; f,g\in \mathfrak{h}$ real valued obeying the Weyl relations 
$W(f,g)^\ast=W(-f,-g)$, $W(f,g) W(f',g')=W(f',g')\; W(f,g)\exp(i[<f,g'>-<f',g>])$ and 
a Fock state $\omega(W(f,g))=\exp(-\frac{1}{4}[||f||^2+||g||^2])$ the GNS representation of
the Weyl algebra $\mathfrak{A}$ generated by the Weyl elements given by the GNS data ${\cal H},\;\rho,\;
\Omega$ of 
$\omega$ is irreducible. Equivalently, $\omega$ is pure or every vector in the GNS Hilbert
space is cyclic \cite{14}. 

The Weyl elements are by definition unitary operators, thus bounded. The Fock representation 
of $\mathfrak{A}$ is strongly continuous for the 1-parameter unitary groups 
$W_t(f,g):=W(tf,tg)$ for any $f,g\in \mathfrak{h}$ real valued and thus we have access to the self-adjoint 
unbounded operators 
$<f,\hat{e}>=s-\lim_{t\to 0}\frac{1}{it} [W_t(f,0)-1],\; <g,\hat{P}>=s-\lim_{t\to 0}\frac{1}{it} [W_t(0,0g)-1]$
with domain of self-adjointness given by elements of $\cal H$ where the strong limit is defined.
This domain includes the span of vectors of the form $W(F,G)\Omega,\; F,G\in \mathfrak{h}$ real valued
and in fact forms an invariant domain of analytic vectors \cite{46}. We thus also have access 
to annihilation and creation operators $a[z],\;a[z]^\ast$ defined for complex valued 
$z=(f-ig)/\sqrt{2}\in \mathfrak{h}$ by 
$a[z]+a[z]^\ast=<f,\hat{e}>+<g,\hat{P}>$. One may check that on the span of the $W(F,G)\Omega$ we have 
the non-trivial 
CCR $[a[z],a[z']^\ast]=<z,z'>\; 1_{{\cal H}}$ and vacuum property $a[z]\Omega=0$. 
The connection between these concept is given by the formula
\be \label{7.1}
W(f,g)=e^{-||z||^2/2}\; e^{i\;a[z]^\ast}\; e^{i\;a[z]},\; z=\frac{1}{\sqrt{2}}[f-ig]      
\ee
as one may check by verifying that (\ref{7.1}) obeys the Weyl relations
(using the Baker Campbell Hausdorff formula on analytic vectors for $\mathfrak{A}$ 
\cite{46}). 

It follows that 
\be \label{7.2}
W(f,g)\Omega=\Omega_{iz},\; \Omega_\zeta:=e^{-||\zeta||^2/2}\; e^{a[\zeta]^\ast}\Omega
\ee
is a {\it coherent state}, i.e. a minimal uncertainty vector for the $<f,e>,\;<g,P>$ 
and eigenstate of the annihilation operator $a[z']\;\Omega_\zeta=<z',\zeta>\; \Omega_\zeta$

Consider now polynomials in creation and annihilation operators applied to a fixed 
coherent state $\Omega_\zeta$. By normal ordering using the CCR and exploiting 
the eigenstate property of the coherent state, we can restrict attention to the span of the states
$\Omega_\zeta$ and
\be \label{7.3} 
a[z_1]^\ast\;..a[z_N]^\ast\; \Omega_\zeta
\ee
These are normalisable elements of $\cal H$ and thus can be approximated arbitrarily closely 
by elements in the span of the $W[F,G]\Omega$. In fact, consider for $z=(f-ig)/\sqrt{2}$
\be \label{7.4}
a_t(z)^\ast:=\frac{1}{2it}\{([W(tf,0)-1)+i(W(0,tf)-1)]-i[W(tg,0)-1)+i(W(0,tg)-1)]\}
\ee
then using strong continuity it is not difficult to see that 
\be \label{7.5}
a_{t_1}[z_1]^\ast...a_{t_N}[z_N]^\ast \Omega_\zeta
\ee
strongly converges to (\ref{7.3}) as $t_1,..,t_N\to 0$. Thus (\ref{7.5}) provides a 
concrete approximant of (\ref{7.3}) in the span of the $W[F,G]\Omega$ since 
using the Weyl relations and (\ref{7.2}) we can write (\ref{7.5}) as a concrete element 
in that span. 

Conversely, given $\Omega_\zeta$ consider 
\be \label{7.6}
\Omega^N_\zeta:=e^{-||\zeta||^2/2}\;\sum_{n=0}^N\;\frac{1}{n!}\; (a[\zeta]^\ast)^n \Omega
\ee
Then it is not difficult to see that $\Omega^N_\zeta$ converges strongly to $\Omega_\zeta$
as $N\to \infty$. Now for any $\psi\in {\cal H}$ and $\epsilon>0$ we find a finite linear 
combination $\psi_0=\sum_{k=1}^M c_k \Omega_{\zeta_k}$ such that $||\psi-\psi_0||<\epsilon/2$ 
and for every $\zeta_k$ we find $N_k$ such that $||\Omega^{N_k}_{\zeta_k}-\Omega_{\zeta_k}||<\frac{\epsilon}{2M|c_k|}$.
Accordingly $||\psi_0-\psi_1||<\epsilon/2$ with $\psi_1=\sum_{k=1}^M \;c_k\;\Omega^{N_k}_{\zeta_k}$. 
whence $||\psi-\psi_1||\le ||\psi-\psi_0||+||\psi_0-\psi_1\|<\epsilon$. This shows that the 
span of the $a[z_1]^\ast..a[z_N]^\ast \Omega$ is also dense. By decomposing $z_k=(f_k-ig_k)/\sqrt{2}$ it follows that 
also the span of the $a[z_1]^\ast..a[z_N]^\ast \Omega$ with real valued $z_k$ is dense. 

Now consider again fixed $\zeta$. By irreducibility, $\Omega_\zeta$ is cyclic for $\mathfrak{A}$. Hence
any $\psi\in {\cal H}$ can be approximated by a finite linear combination of elements of the form 
$W(f,g)\Omega_\zeta$. Using (\ref{7.1}) and the eigenvalue property, this is the same as the span of the 
$\Omega_{z,\zeta}:=e^{a[z]^\ast}\; \Omega_\zeta$. Consider 
\be \label{7.7}
\Omega^N_{z,\zeta}:=\sum_{n=0}^N\;\frac{1}{n!}\;(a[z]^\ast)^n\; \Omega_\zeta
\ee
It is not difficult to show that $s-\lim_{N\to\infty}\;\Omega^N_{z,\zeta}=\Omega_{z,\zeta}$.
Thus, also the span of $\Omega_{\zeta}$ and the $a[z_1]^\ast..a[z_N]^\ast\Omega_\zeta$ is dense and again
by decomposing into real and imaginary part the span of those vectors with $z_1,..,z_N$ real valued is also 
dense. Finally again by strong continuity we can approximate these vectors arbitrarily closely by 
the vectors 
\be \label{7.8}
\frac{e^{t_1\;a[z_1]^\ast}-1}{t_1}..\frac{e^{t_N\;a[z_N]^\ast}-1}{t_N}\;\Omega_\zeta
\ee
with $z_1,..,z_N$ real valued. Due to (\ref{7.1}) we have
$e^{ta[z]^\ast}=e^{||z||^2/2}\;W(0,g)\; e^{-it a[z]}$ with $z=g/\sqrt{2}$ real valued  
hence (\ref{7.8}) is in the span of the vectors of the form 
$W(0,g)\Omega_\zeta$. Thus we have established:
\begin{Lemma} \label{la7.1} ~\\
Let $\zeta\in \mathfrak{h}$ be a fixed complex valued element. Then 
\be \label{7.9}
{\cal D}_\zeta:={\sf span}\{W(0,g)\Omega_\zeta;\; g=g\ast\in \mathfrak{h}\},\;\;
{\cal D}'_\zeta:={\sf span}\{\Omega_\zeta,\;a[f_1]^\ast..a[f_N]^\ast\Omega_\zeta;\; f_k=f_k^\ast\in \mathfrak{h},\;
N\in \mathbb{N}\}
\ee
are dense in $\cal H$.
\end{Lemma}
The significance of these sets will be that they define dense domains of geometrical quadratic forms.\\
\\
Now consider some operator $K$ on the 1-particle Hilbert space $\mathfrak{h}$.
We consider the object
\be \label{7.10}
\Omega_K:= e^{\frac{1}{2}\;<a,K\cdot a^\ast>}\;\Omega,\;\; <f,K\; g>=
\int\; d^Dx\;\int\;d^Dy\; [f^J_j(x)]^\ast\; K_{JL}^{jl}(x,y)\; g^L_l(y) 
\ee
We see that the integral kernel of $K$ can be chosen symmetric under $x,J,j\leftrightarrow y,L,l$.
When it is real valued then $K$ is symmetric. If it is bounded, then it is self-adjoint. 
We will assume that its spectrum is pure point. Then $\mathfrak{h}$ has an orthonormal basis
consisting of eigenvectors $b_n$ which can be chosen real valued (decompose into real 
and imaginary part) of $K$ with eigenvalues $k_n$
When expanding the exponential function, we see that it is a superposition of $N-$particle 
states with even particle number. We call $\Omega_K$ a {\it squeezed state} when $||\Omega_K||<\infty$ 
otherwise a {\it squeezed distribution}. 

We note that for any $g\in \mathfrak{h},\; \psi\in {\cal H}$
\ba\label{7.11}
&& <\Omega_K,<g,\hat{P}>\;\psi>=<<g,\hat{P}>\;\Omega_K,\;\psi>
=-\frac{i}{\sqrt{2}}<[<g,a>-<g,a^\ast>]\;e^{\frac{1}{2}<a,K\; a^\ast>}\Omega,\;\psi>
\nonumber\\
&=& -\frac{i}{\sqrt{2}}<<[K-1]g,a^\ast>]\;\Omega_K,\;\psi>
\ea
It follows that for the choice $K=1_{\mathfrak{h}}$ (bounded and self-adjoint with pure 
point spectrum), $\Omega_K$ is {\it annihilated by the momentum operator}. A similar 
computation shows that for $K=-1_{\mathfrak{h}}$ the object is $\Omega_K$ is {\it annihilated by 
the D-bein operator $\hat{e}$}. However, in both cases $\Omega_K$ is not normalisable as we will 
now show which means that zero eigenstates of these operators can be understood within 
the Fock space as {\it generalised eigenstates}. This is interesting because it offers 
the possibility to connect the Fock representation to the Narnhofer-Thirring representation
in which $\hat{e}$ or $\hat{P}$ rather than $a$ annihilates the GNS vacuum. 

To compute $||\Omega_K||$ we we define the mode annhilation and creation operators 
$a_n=<b_n,a>,\; a_n^\ast$ that obey the CCR $[a_m,a_n]=\delta_{m,n}$. 
Then $<a,K a^\ast>=\sum_n\; k_n\; [a_n^\ast]^2$ and the computation factorises
\be \label{7.12}
||\Omega_K^2||=\prod_n\; ||e^{\frac{k_n}{2}\; [a_n^\ast]^2}\;\Omega||^2=
\prod_n s(k_n)
\ee
where 
\be \label{7.13}
s(k)=\sum_{m=0}^\infty\; \frac{k^{2m}\;(2m)!}{2^{2m}\;[m!]^2}=[1-k^2]^{-1/2}
\ee
The last equality is only true when $k^2<1$ as indeed $k^2=1$ is the radius 
of convergence of the series. Thus, the first condition for (\ref{7.12}) to 
be finite is that $k_n^2<1$ for all $n$. Next the infinite product in (\ref{7.12})
converges iff its logarithm
\be \label{7.14}
-\frac{1}{2}\sum_n \ln(1-k_n^2)
\ee
converges. Thus $k_n^2$ must converge to zero. 
Using that $-\ln(1-x)\le x+x^2,\;0\le x<1/2$ we see that for $n>n_0$ where $k_n^2\le 1/2$ for 
$n>n_0$ the sum can be estimated from above by 
%f(x)=\ln(1-x)+x+x^2, f'(x)=-1/(1-x)+1+2x=-x/(1-x)+2x=x(2[1-x]-1)/1-x=x(1-2x)/1-x 
\be \label{7.15}
\frac{1}{2}[-\sum_{n\le n_0}\ln(1-k_n^2)+\sum_{n>n_0} k_n^2(1+k_n^2)]
\le \frac{1}{2}[-\sum_{n\le n_0}\ln(1-k_n^2)+\frac{3}{2}\sum_{n>n_0} k_n^2
\ee 
which converges iff 
\be \label{7.16}
\sum_n k_n^2 ={\sf Tr}(K^2)<\infty 
\ee
i.e. $K$ must be a Hilbert-Schmidt operator which also grants that 
$\mathfrak{h}$ has an orthonormal basis diagonalising $K$. 

Note that HS operators are 
in particular bounded. Let $k^2=\sup \{k_n^2,\; n\in \mathbb{N}\}$.
Then 
\be \label{7.17}
||K\psi||^2=\sum_n\; k_n^2\; |<b_n,\psi>|^2\le k^2\sum_n\;|<b_n,\psi>|^2\le k^2\; ||\psi||^2,\;
\ee
which shows that $||K||=|k|$ because $||K b_n||=|k_n|$ and we can find a subsequence of the 
$b_n$ such that $k_n\to k$. Thus we have 
\begin{Lemma} \label{la7.2} ~\\
$||\Omega_K||<\infty$ iff $||K||<1$ and ${\sf Tr}(K^2)<\infty$.
\end{Lemma}
The choice $K=\pm 1$ violates both conditions. However, consider the operator
\be \label{7.18}
K^\epsilon=\pm\sum_n \; k_n^\epsilon \; b_n\; <b_n,.>,\;\;k^\epsilon_N:=\frac{1}{[1+\epsilon]\sqrt{1+[\epsilon n]^2}}
\ee
Then for any $\epsilon>0$ we have $||K^\epsilon||=(1+\epsilon)^{-1}<1$ and 
\be \label{7.19}
{\sf Tr}(K^2)=[1+\epsilon]^{-2}\sum_{n=1}^\infty\;[1+(\epsilon n)^2]^{-1}
\le [1+\epsilon]^{-2}\int_0^\infty\; dx\; [1+(\epsilon x)^2]^{-1}=\frac{c}{\epsilon[1+\epsilon^2]}
\ee
with $c=\int_0^\infty\; dx\; [1+x^2]^{-1}$. Thus $K^\epsilon\to 1_{\mathfrak{h}}$ uniformly 
with respect to $\mathfrak{h}$ as $\epsilon\to 0+$ but for any $\epsilon >0$ we have 
$||\Omega_{K^\epsilon}||_{{\cal H}}<\infty$. Hence the Narnhofer-Thirring ``distribution''
$\Omega_{\pm 1_{\mathfrak{h}}}$ can be understood as a limit of the states 
$\Omega_{K^\epsilon}$ in the Fock space where the convergence is in the sense 
of linear forms on ${\cal D}'_0$, the span of the excitations of the Fock vacuum $\Omega$
by polynomials of excitation operators $a[f]^\ast$. Indeed
\be \label{7.20}
<\Omega_K,a[f_1]^\ast..a[f_N]^\ast\;\Omega>=
<\Omega,(a[f_1]^\ast+a[K\cdot f_1])..(a[f_N]^\ast+a[K\cdot f_N])\;\Omega>     
\ee
thus $\Omega_K$ defines a well defined linear form on ${\cal D}'_0$ whether or not it 
it is normalisable and (\ref{7.20}) is continuous under $K^\epsilon\to \pm 1_{\mathfrak{h}}$.
This is similar to the way that tempered distributions are weak limits of Schwartz functions
except that we have not specified a topology on ${\cal D}'_0$ finer than the subspace topology 
inherited from $\cal H$ (also we replace $\Omega_K$ by $\Omega_K^N\in {\cal D}'_0$
given by the first $N$ terms in the expansion of the exponential function). 
Doing so would enable us to define $\Omega_K$ as a continuous 
linear functional on ${\cal D}'_0$ (it cannot be with respect to the Hilbert space topology
because of the Riesz representation theorem) but this would take us too far apart at this point.

\section{Geometrical quadratic forms}
\label{s8}

Consider a $d-$dimensional, compact, embedded submanifold $S\subset\sigma,\; d=1,..,D$ with  
embedding $E_S:\;s\subset \mathbb{R}^d\to S;\;y\to x=E_S(y)$. Its ``volume'' is defined by 
\be \label{8.1}
{\sf Vol}[S]=\int_s\; d^dy\; \sqrt{\det(E_S^\ast q)},\;\; q_{ab}=e^j_a e^k_b\delta_{jk}   
\ee
This is an example of a highly non-polynomial function when expressed in terms of our 
annihilation and creation operator $a,a^\ast$. It is therefore unclear how to quantise 
it and it is often claimed to be impossible in Fock representations and thus provides 
one of the motivations for the LQG representation where such quantisations exist as 
self-adjoint operators with pure point spectrum. Recall that 
\be \label{8.2}
e_a^j=h_a^I\;|\det{h}|^{-1/2}\; \hat{e}_I^j,\; h_a^I=[\frac{\hat{X}^I}{|\det(\hat{e})|^{1/D}}]_{,a}
\ee
in the case that we do not insist of having a polynomial version of the Hamiltonian constraint.
Thus (\ref{8.1}) is the integral over the points $x\in S$ of a function which depends 
rather non-polynomially on $\hat{e}_I^j, \hat{X}^I$ and their first spatial derivatives
$\hat{e}_{I,a}^j, \hat{X}^I_{,a}$   

More generally we may be confronted with integrals over $x$ of functions $F$ which depend on 
$\hat{e}_I^j,\;\hat{P}^I_j,\hat{X}^I,\hat{Y}_I$ and their first spatial derivatives at that 
point $x$ up to some order $M$. We need to write these in a meaningful way in terms 
of the the annihilation and creation operators. This reminds of a similar problem in 
quantum mechanics. Consider the phase space given by the cotangent bundle $T^\ast \mathbb{R}^n,\;
n<\infty$. Given a phase space function $F=F(q,p)$ one defines the {\it Weyl quantisation} \cite{26} of 
$F$ using the Fourier transform by 
\be \label{8.3}
W_F:=\int_{\mathbb{R}^{2n}} \; \frac{d^n f\;d^n g}{[2\pi]^{2n}}\; W(f,g)\; \hat{F}(f,g),\;
\hat{F}(f,g):=\int\; d^n x\; d^n y\; F(x,y)\; e^{-i[f\cdot x+g\cdot y]},\;
W(f,g)=e^{i[f\cdot q+g\cdot p]} 
\ee
This is related to deformation quantisation \cite{27} and the Moyal product
$W_F\; W_G=W_{F\ast G}$ where $F\ast G-G\ast F=i\hbar\{F,G\}+O(\hbar^2)$. Thus one could be tempted 
to apply Weyl quantisation to the above task. The immediate problem is that in field 
theory we have $n=\infty$ and (\ref{8.3}) is ill defined for $n=\infty$. However, 
we note that what we need is a quantisation prescription for a function which depends 
on a {\it finite set of variables} $v^A, A=1,..,n$ where each $v^A$ stands for 
one of the partial derivatives, of order $0,1,..,M$ of the fields or their conjugate momenta 
{\it at fixed x}. Thus we consider the quantisation prescription 
\be \label{8.4}
F(v)=
\int_{\mathbb{R}^{n}} \; \frac{d^n z}{[2\pi]^n}\; :\;e^{i z_A\;v^A}\; \hat{F}(z)\; :,\;\,
\hat{F}(z):=\int\; d^n y\; F(y)\; e^{-i\; z_A\; y^A}
\ee
Thus we deviate from Weyl quantisation in three important aspects which is forced on us
in field theory: 1. infinite dimensional Fourier transforms are avoided by transforming 
pointwise on $\sigma$, 2. as a price to pay we have to assign an independent Fourier variable   
for each derivative as well and 3. as operator valued distributions cannot be multiplied 
we need to normal order the resulting expression in order to obtain at least a quadratic form.

It may even happen that $F$ appears multiplied by similar non-polynomial functions and/or polynomials. In that 
case we apply (\ref{8.3}) to each of these functions and at the end normal order the 
whole product. This is easily done decomposing $v^A$ into (derivatives of) annihilation and
creation operators and then using e.g. $:e^{i[z\partial^k a+z'\partial^l a^\ast]}:=
e^{i\;z'\partial^l a^\ast}\;e^{i\;z\partial^k a}$. In this way one approach 
to write (\ref{8.1}) as a quadratic form is to write it as 
\be \label{8.5}
{\sf Vol}[S]=\int_s\; d^d y\; 
:F(\hat{e},\partial\hat{e},\hat{X},\partial \hat{X})(E_S(y)):
\ee
where 
\be \label{8.6}
:F:\; =\; \int\; \frac{d^n z}{[2\pi]^n}\;:e^{i[
z^I_j \hat{e}_I^j+z^{Ia}_j\hat{e}_{I,a}^j+ 
z_I \hat{X}^I+z^{a}_I\hat{X}^I_{,a}]}(E_S(y)):
\hat{F}(z)
\ee
$n$ is the number of the variables $z$ on which $F$ depends and 
and $\hat{F}$ is the Fourier transform of $F$. The operator valued distributions
appearing in the exponent all are evaluated at the same point $x=E_S(y)$. 

This procedure potentially delivers well defined quadratic forms on a suitably 
defined dense domain, depending on the 
existence of the Fourier transform $\hat{F}$. To see what candidates for such a domain 
could be note that $<\Omega,\; :\exp(.):\;\Omega>=1$ so that $<\Omega, :F:\Omega>
\propto \int\; d^nz\; \hat{F}(z)\propto F(0)$. Thus there is no chance to define such objects 
as quadratic forms in the manner proposed on the domain ${\cal D}'_0$ (finite linear combinations 
of particle number eigenstates) when $F$ diverges as one or several of the $z$ approach zero. 
Accordingly, from (\ref{8.2}) we see that ${\cal D}'_0$ is not suitable as a domain for (\ref{8.1}).
More generally, if we choose to write the Hamiltonian constraint in non-polynomial form,
then we are confronted with inverse powers of $\det(e)$ and thus face the same problem 
in defining those on the domain ${\cal D}'_0$. We are thus naturally lead to consider 
different domains. 

A natural choice of domain follows from the following consideration: The above problems with 
the zeroes of $F$ do not arise in the classical theory where we only allowed $e_a^j,X^I$ such that 
$\det(e),\det(h), h_a^I=X^I_{,a}$ are nowhere vanishing. Accordingly we are lead to consider 
domains based on semiclassical states. Indeed, for any coherent state $\Omega_\zeta$ we have 
by construction
\be \label{8.7}      
<\Omega_\zeta,\;:F:\;\Omega_\zeta>=F(\hat{e}_\zeta,\partial\hat{e}_\zeta,\hat{X}_\zeta,\partial \hat{X}_\zeta)(x)
\ee
where 
\be \label{8.8} 
(\hat{e}_I^j)_\zeta(x)=<\Omega_\zeta,\;\hat{e}_I^j(x)\;\Omega_\zeta>=\frac{1}{\sqrt{2}}(\zeta_I^j(x)+\bar{\zeta}_I^j(x))
\ee
is the expectation value of $\hat{e}_I^j(x)$ at which $\Omega_\zeta$  
is concentrated and similar for the other variables. Hence, when those 
expectation values correspond to non-degenerate $\hat{e},h$ then (\ref{8.7})
is well defined provided the Fourier transform $\hat{F}$ exists. We thus 
aim for a domain based on $\zeta$ such that $\hat{e}_\zeta,h_\zeta$ are non-degenerate. 

Now consider 
\be \label{8.9}
<\Omega_\zeta,\; :F:\; \Omega_{\zeta'}>
\ee
The integral defining $:F:$ contains e.g. the Weyl factors of the form 
\be \label{8.10}
:e^{i z^I_j \hat{e}_I^j(x)}:\;=\;e^{i z^I_j a_I^j(x)^\ast/\sqrt{2}}\; e^{i z^I_j a_I^j(x)/\sqrt{2}}\; 
\ee
for which we obtain 
\be \label{8.11}
<\Omega_\zeta,\; :e^{i z^I_j \hat{e}_I^j(x)}:\;\Omega_{\zeta'}>=e^{i\; z^I_j[(\zeta')_I^j(x)+\bar{\zeta}_I^j(x)]}\;
<\Omega_\zeta,\Omega_{\zeta'}>
\ee
Thus the integral over $z$ will generically diverge unless $\zeta,\zeta'$ have identical imaginary part. 
Hence a suitable candidate domain consists of the span of the states $W(0,g)\Omega_\zeta$ with $\zeta$ such 
that $\hat{e},h$ have non-degenerate expectation values. This is precisely on of the domains  
${\cal D}_\zeta$
which we have shown to be dense in the previous section. By 
(\ref{7.2}) this domain consists of the span of coherent states $\Omega_{\zeta'}$ with fixed imaginary 
part $\Im(\zeta')=\Im(\zeta)$.

We now must deal with the existence of the Fourier transform. Consider for instance the 
function $F=|\det(\hat{e})|$. In the gauge $X^I=\delta^I_a x^a$ it coincides with $|\det(e)|$,
thus the integral of $F$ over a D-dimensional region is an interesting geometrical quadratic 
form because in that gauge it measures the $D-$volume of $S$ with respect to $e$. 
The problem is that the Fourier transform 
\be \label{8.12}
\hat{F}(z)=\int\; d^{D^2}k\; e^{-i k_I^j z^I_j}\; |\det(k)|
\ee
certainly does not exist as a function, at best it defines a Schwartz distribution. To
improve this, we modify the procedure laid out so far and start from the identity
\be \label{8.13}
I_{F,r}(\hat{e}):=|\det(\hat{e})|^{-(D+r)}=c_r\;\int\; d^{D^2}z\; |\det(z)|^r\; F(z\cdot \hat{e}),\;
c_r^{-1}= \int\; d^{D^2}z\; |\det(z)|^r\; F(z),     
\ee
for any $r\in \mathbb{R}_+$ and any function $F:\; \mathbb{R}^{D^2}\to \mathbb{C}$. 
The integrals converge provided that $c_r$ does which will be the case whenever $F$ 
decays faster than any power of any of the $z^I_j$. For instance 
$F=\exp(-{\sf Tr}(z^T z))$ could be a Gaussian. This suggests 
to consider the identity
\be \label{8.14}
|\det(\hat{e})|=[\det(\hat{e})]^{2k} \;I_{F,2k-D-1}(\hat{e}) 
%D+r=2k-1,r=2k-D-1 
\ee
for any $k\ge (D+1)/2$. The idea is now to 
express 
\be \label{8.15}
F(z\cdot \hat{e})=\int\; \frac{d^{D^2}p}{(2\pi)^{D^2}}\; e^{i{\sf Tr}(p\cdot z\cdot \hat{e})} \; \hat{F}(p)
\ee
as a Fourier integral and to define 
\be \label{8.16}
:|\det(e)|: \; = \;c_r \;\int\, d^{D^2}z \; |\det(z)|^r\;\int\; \frac{d^{D^2}k}{(2\pi)^{D^2}}\; 
\hat{F}(p)\;
:\; e^{i\;p\cdot z\cdot \hat{e}} \; [\det(\hat{e})]^{2k} \;:
\ee
with $r=2k-D-1$ and $k$ the smallest integer such that $k\ge (D+1)/2$. Then the previous 
investigations still apply and there is no question that $\hat{F}$ exists. 
Accordingly, we can define 
\be \label{8.17}
\widehat{{\sf Vol}}[S]=\int\; d^Dy\; |\det((\partial E_S(y)/\partial y))|\;
:\;|\det(\hat{e})(E_S(y))|\; :
\ee
as a quadratic form densely defined on ${\cal D}_\zeta$ or  ${\cal D}'_\zeta$
for any $\zeta$ such that $\hat{e}_\zeta$ is non-degenerate. In particular 
by construction the expectation value of (\ref{8.18}) with respect to $\Omega_\zeta$ 
coincides with the classical value of (\ref{8.17}) at $\hat{e}=\hat{e}_\zeta$. 

How would one quantise (\ref{8.1}) itself? We have 
\ba \label{8.18}
&& |\det(e)|=|\det(\hat{e})|\;|\det(h)|^{1-D/2},\;
\nonumber\\
&& \det(h)
=\det(\hat{h})\;(|\det(\hat{e})|^{-1/D}+(|\det(\hat{e})|^{-1/D})_{,a} \hat{h}^a_I \hat{X}^I)
\nonumber\\
&=&|\det(\hat{e})|^{-1/D}\;\det(\hat{e})^{-1}\;[\det(\hat{h})\det(\hat{e})
-\frac{1}{D} \det(\hat{h})\hat{h}^a_I \hat{X}^I (\det(\hat{e}))_{,a}]
\ea
Thus we can write $|\det(e)|=F(\hat{e},\partial\hat{e},\hat{X},\partial\hat{X})$ and 
then proceed in principle as above. In this case the Fourier transform of $F$ 
involves $D^2+D^3+D+D^2$ variables and it is likely to exist only as a distribution
which is sufficient in order to compute coherent state expectation values.
One may be able to improve this by using additional identities along the lines of 
(\ref{8.13}). We keep this for future research.

Instead we now consider geometrical quadratic forms for $d<D$ dimensional $S$. 
For instance 
\be \label{8.19}
\widehat{{\sf Ar}}(S)=\int_s\; d^2y\; \sqrt{[n_I(y) D^I_j(y)]^2},\;
n_I(y)=\delta_I^a \epsilon_{abc} E^b_{S,\alpha}(y)E^b_{S,\beta}(y)\epsilon^{\alpha\beta}/2,\;
D^I_j(y)=\epsilon^{IJK}\epsilon_{jkl} (\hat{e}_J^k\hat{e}_Kl)(E_S(y))/2
\ee
%
%E^a_I E^B_J q_{ab}= (e^j_a E^a_I) (e^j_b E^b_J)=m_{IJ}
%det(m)=m_{11} m_{22}-m_{12}^2 
%=1/2\epsilon^{IJ}\epsilon^{KL} m_{IK} m_{JL}
%=(\epsilon^{IJ} E^a_I E^c_J)(\epsilon^{KL} E^b_K E^d_L) (e^j_a e^k_c) (e^j_b e^k_d)/2
%=n_e n_f (\epsilon^{eac} e_a^j e_c^k)(\epsilon^{fbd} e_b^j e_d^k)/2
%= (n_a D^a_j)^2 = (n_I D^I_j)^2
%n_a=\epsilon^{IJ}\epsilon_{abc} E^b_I E^c_J/2, 
%D^a_j=\epsilon^{abc}\epsilon_{jkl} e_b^k e_c^l/2= h_b^M h_c^N 
%\epsilon^{abc}\epsilon_{jkl} \hat{e}_M^k \hat{e}_N^l/2/det(h) 
%=h^a_P D^P_j
%det(m)=(n_P D^P_j)^2, n_P=h^a_P n_a
%D^P_j=\epsilon^{PMN}\epsilon_{jkl} e_M^k e_N^l/2= 
which coincides with 
the area $d=2$ of $S$ in $D=3$ in the gauge $X^I=\delta^I_a x^a$ with respect to $e$. An efficient way 
to quantise (\ref{8.19}) is to introduce $u^I(y), v^I(y)$ such that 
%$\delta_{IJ} u^I v^J=0$ and 
$n_I=\epsilon_{IJK} u^J v^K$ whence 
$n_I D^I_j=\epsilon_{jkl} (u^I \hat{e}_I^k) ((v^J \hat{e}_J^l)$. Then 
(\ref{8.19}) just depends on the variables $p^j(y)= u^I(y) \hat{e}_I^j(E_S(y)), 
q^j= v^I(y) \hat{e}_I^j(E_S(y))$
and what we need is the Fourier transform 
\be \label{8.20}
\hat{F}_r(\vec{a},\vec{b})=\int \;d^3p \;d^3q \; e^{-i[a^j p_j+b^j q_j]}\; ||\vec{p}\times \vec{q}||^{-r}
\ee
for $r>0$ such that 
\be \label{8.21}
\widehat{{\sf Ar}}(S)=\int_s\; d^2y\; 
\int\; \frac{d^{3M} a}{(2\pi)^{3M}}\; \frac{d^{3M} b}{(2\pi)^{3M}}\;\prod_{k=1}^M\; \hat{F}_{r_k}(\vec{a}_k,\vec{b}_k) \;
:\; || \vec{p}(y)\times\vec{q}(y)||^{2N}\; e^{i [\sum_k\vec{a}_k\cdot \vec{p}(y)+\sum_k\vec{b}_k)\cdot \vec{q}(y)]}\;: 
\ee
where $\sum_{k=1}^M r_k=2N-1$, e.g. $r_k=r=(2N-1)/M$ with $N+1,M\ge 1$.
The integral (\ref{8.20}) is superficially diverging for large $\vec{p}, \vec{q}$ at 
the natural values $N=0,M=1$ which is why we have considered this generalisation.
The integrand now involves a polynomial and an exponential of $p^j(y),q^j(y)$ and thus 
is easily normal ordered.

To compute (\ref{8.20}) we proceed as follows. Ignoring the singular value $\vec{p}=0$ we define 
$\vec{n}_3(p)=\vec{p}/||\vec{p}||$ and $\vec{n}_1(p)=\partial_\theta \vec{n}_3(p)$ and 
$\sin(\theta)\vec{n}_2(p)=\partial_\phi \vec{n}_3(p)$ with respect to polar coordinates \\
$\vec{n}(p)=(\sin(\theta)\cos(\phi),\sin(\theta)\sin(\phi),\cos(\theta))$. At fixed 
$\vec{p}$ we denote by $q_j(p),b_j(p)$ the components of $\vec{q},\vec{b}$ with respect 
to this orthonormal basis and rotate the $\vec{q}$ integral accordingly. Then 
\ba \label{8.22}
\hat{F}_r &=& \int\; \frac{d^3p}{||\vec{p}||^r}\; e^{-i\vec{p}\cdot\vec{a}}
\int\; d^3 q\; \frac{e^{-i(\sum_j q_j b_j(p))}}{\sqrt{q_1^2+q_2^2}^r} 
\nonumber\\
&=& 2\pi \int\; \frac{d^3p}{||\vec{p}||^r}\; e^{-i\vec{p}\cdot\vec{a}} \delta(b_3(p))
\int_0^\infty d\rho \rho^{1-r}\int_0^{2\pi} e^{-i\rho\; B(p)\cos(\phi)}
\nonumber\\
&=& (2\pi) \; C_{1-r}\; \int\; \frac{d^3p}{||\vec{p}||^r}\;B(p)^{r-2}\; e^{-i\vec{p}\cdot\vec{a}} \delta(b_3(p),0)
\ea
where $B(p)=\sqrt{b_1(p)^2+b_2(p)^2}$ and 
\be \label{8.23}
C_s:=\int_0^\infty d\rho \rho^s\int_0^{2\pi} e^{-i\rho\;\cos(\phi)}
\ee
is a generalised Bessel integral. Next we introduce $n_j(b)$ in analogy to $n_j(p)$
and write 
\be \label{8.24}   
B(p)=\sqrt{||\vec{b}||^2-(\vec{p}\cdot \vec{b})^2/||\vec{p}||^2}=||\vec{b}||
\sqrt{1-p_3(b)^2/||\vec{p}||^2}
\ee
where we have denoted the components of $\vec{p},\vec{a}$ with respect to $\vec{n}_j(b)$
by $p_j(b),a_j(b)$ respectively. We also have 
\be \label{8.25}
\delta(b_3(p),0)=\delta((\vec{b}\cdot{p})/||\vec{p}||,0)=\delta(p_3(b)\frac{||\vec{b}||}{||\vec{p}||},0)
=\frac{\delta(p_3(b),0)}{|\frac{||\vec{b}||}{||\vec{p}||}(1-\frac{p_3(b)^2}{||\vec{p}||^2})|}
%
%\partial_{p_3} p_3 b/\sqrt{p_3^2+p_1^2+p_^2}=b(1/p-p_3^2/p^3)
\ee
We rotate the $\vec{p}$ coordinate system and carry out the integral over $p_3$
\ba \label{8.26}
\hat{F}_r 
&=& (2\pi) \; C_{1-r}\; ||\vec{b}||^{r-3}\;\int\; 
\frac{d^3p}{||\vec{p}||^{r-1}}\;\sqrt{1-p_3^2/||\vec{p}||^2}^{2-r}\; e^{-i\sum_j p_j a_j(b)} \delta(p_3,0) 
\nonumber\\
&=& (2\pi) \; C_{1-r}\; ||\vec{b}||^{r-3}\;\int\; 
\frac{d^2p}{\sqrt{p_1^2+p_2^2}^{r-1}}\; e^{-i(p_1 a_1(b)+p_2 a_2(b))}  
\nonumber\\
&=& (2\pi) \; C_{1-r}\; ||\vec{b}||^{r-3}\;\int_0^\infty \;\frac{d\rho}{\rho^{r-2}} \int_0^{2\pi}\; d\phi
\; e^{-i\rho A(b)\; \cos(\phi)}  
\nonumber\\
&=& (2\pi) \; C_{1-r}\; \; C_{2-r}||\vec{b}||^{r-3}\;A(b)^{r-3}
%\int_0^\infty \;\frac{d\rho}{\rho^{r-2}} \int_0^{\2\pi}\; d\phi
%\; e^{-i\rho A(b)\; \cos(\phi)}  
\nonumber\\
&=& (2\pi) \; C_{1-r}\; \; C_{2-r}\; ||\vec{a}\times\vec{b}||^{r-3}
\ea
where $A(b)=\sqrt{a_1(b)^2+a_2(b)^2}$. Thus $\hat{F}_r$ exists provided that $C_{1-r}, C_{2-r}$ do.

The Bessel function is defined as 
\be \label{8.27}
J_n(\rho)=\frac{1}{\pi}\int_0^\pi\; d\phi\; \cos(n\phi-\rho\sin(\phi))
\ee
which can be extended to complex $n$. 
It follows 
\be \label{8.28}
J_0(\rho)=\frac{1}{2\pi}\int_0^{2\pi}\; d\phi\; e^{-i\rho\cos(\phi)}
%=\int_0^\pi [e^{-i\rho cos(\phi)}+e^{-i\rho cos(\pi+\phi)}]/(2\pi)
%=\int_0^\pi \cos(\rho cos(\phi)}/\pi
%=\int_0^{\pi/2} (\cos(\rho cos(\pi/2+\phi)}+\int_{\pi/2}^\pi \cos(\rho \cos(\phi-\pi/2))/\pi
%=\int_0^{\pi/2} (\cos(\rho sin(\phi)}+\int_{\pi/2}^\pi \cos(-\rho \sin(\phi))/\pi
%=J_0(\rho)
\ee
Integrals of Bessel functions are studied in \cite{48}. Of interest are the integrals 
\be \label{8.29}
I_{m,n}:=\int_0^\infty\; d\rho\; \rho^m\; J_n(\rho)=2^m\; 
\frac{\Gamma(\frac{1}{2}(n+m+1))}{\Gamma(\frac{1}{2}(n-m+1))}
\ee
which holds for $\Re(n+m)>1, \; \Re(m)<1/2$ and $\Gamma$ is the Gamma function which 
is defined everywhere in the complex plane except at the non-positive integers. 
In our case we are interested 
in $n=0$ which confines the  validity of (\ref{8.29}) to $-1<\Re(m)<\frac{1}{2}$. 
In addition we must avoid the right hand side to diverge which would happen 
for $(1+m)/2=0,-1,-2,..$ or to become zero which would happen for 
$(1-m)/2=0,-1,-2,..$. Thus we must avoid $m$ to be any odd integer. Among the 
real values of $m$ this leaves the range $-1 < m < \frac{1}{2}$.

The relation to our application is 
$C_{1-r}=2\pi I_{1-r,0},\; C_{2-r}=2\pi I_{2-r,0}$ hence in our application 
$m=1-r,2-r$. Accordingly the allowed range is $\frac{3}{2} <r< 2$. 
%-1<1-r<1/2,-1< 2-r< 1/2 ==> -2<-r<-1/2, -3<-r<-3/2 ==> 1/2 < r < 2, 3/2 < r < 3
Our values of $r$ take the form $r=(2N-1)/M,\;M,N\ge 1$. The minimal value of 
$N,M$ for which $r$ is in the allowed range is $N=3,M=3$ for which 
$r=5/3$. Summarising 
\be \label{8.30}
\hat{F}_{5/3} 
%=(2\pi)^3 \;I_{1-5/3,0}\; I_1{2-5/3}\; ||\vec{a}\times \vec{b}||^{5/3-3}
=(2\pi)^3 \; 2^{-1/3}\; 
\frac{\Gamma(-\frac{1}{3})}{\Gamma(\frac{4}{3})}
\frac{\Gamma(\frac{2}{3})}{\Gamma(\frac{1}{3})}
||\vec{a}\times \vec{b}||^{-4/3}
=: c\; ||\vec{a}\times \vec{b}||^{-4/3}
%\Gamma(-1/3)=(-3)(-1/3)\Gamma(-1/3)=-3\Gamma(2/3)
%\Gamma(z)=\int_0^\infty dt e^{-t} t^{z-1}, \Re(z)>0 and then extend by $z\Gamma(z)=\Gamma(z+1) 
\ee
and a concrete quantisation of (\ref{8.19}) is given by 
\be \label{8.31}
\widehat{{\sf Ar}}(S)= c^3\;
\int_s\; d^2y\; 
\int\; \frac{d^{9} a}{(2\pi)^{9}}\; \frac{d^{9} b}{(2\pi)^{9}}\;\prod_{k=1}^3\; ||\vec{a}_k\times \vec{b}_k||^{-4/3} \;
:\; || \vec{p}(y)\times\vec{q}(y)||^{6}\; e^{i [\sum_k\vec{a}_k\cdot \vec{p}(y)+\sum_k\vec{b}_k\cdot \vec{q}(y)]}\;: 
\ee

Finally we consider the case $d=1$ in $D=3$ and the length functional
\be \label{8.32}
\widehat{{\sf Len}}[S]:=\int_s\; dy\; \sqrt{\sum_j [\hat{e}^j_S(y)]^2},\;
\hat{e}^j_S(y):=\hat{e}_I^j(E_S(y)) \delta^I_a \frac{d E^a_S(y)}{dy}
\ee
which in the gauge $X^I=\delta^I_a x^a$ is the length of the curve $S$ with respect to $e$.
The quantisation of (\ref{8.32}) is accomplished by 
\be \label{8.33}
\widehat{{\sf Len}}[S]:=\int_s\; dy\; \int\, \frac{d^3 a}{(2\pi)^3}\; \hat{F}(\vec{a})\;
:\;\sum_j [\hat{e}^j_S(y)]^2\; e^{i\sum_j a_j e^j_S(y)}
\; :
\ee
where we have the well known integral, proportional to the Green function of the Laplace operator
\be \label{8.34}
\hat{F}(\vec{a})=\int \; d^3p\; e^{-i \vec{a}\cdot \vec{p}}\; \frac{1}{||\vec{p}||^2} 
% =(2\pi)\int_0^\infty dp \int_{-1}^1 dt e^{-ip a t} 
% =(2\pi)/(-i a) \int_0^\infty dp/p (e^{-ip a}-e^{ipa})
% =(2\pi)/(i a) \int_0^\infty dp/p (e^{ip a}-e^{-ipa}) 
% =(2\pi)/(i a) \lim_{s\to 0+} \int_{\mathbb{r}-[-s,s]} dp/p e^{ip a} % extend to upper half circle, a>0, add subtract
% semi circle of radius s, use Cauchy integral
% =(2\pi)/(i a) \lim_{s\to 0+} \int_0^\pi  [dp/p e^{ip a}]_{p=s e^{i\phi}} 
=\frac{2\pi^2}{||\vec{a}||}  
\ee

We conclude by mentioning that similar techniques work for any $1\le d\le D$ and 
can be used to define inverse volume factors in case that one works with 
non-polynomial Hamiltonian constraints or in a reduced phase space approach 
where the reduced Hamiltonian necessarily is non-polynomial. 

\section{Conclusion}
\label{s9}

In the present paper, which is an expanded version of \cite{28a} we have outlined
in detail how one can use compound fields, using only the experimentally verified 
field content of the gravitational field and the matter fields of the standard 
model, or fundamental scalar fields provided they exist, in order to provide a background independent Fock representation in which 
all constraints of the theory, except for the Hamiltonian constraint, are self-adjoint 
operators. The Hamiltonian constraint itself can be constructed as a symmetric quadratic form
on the same Fock space and there are no LQG like "polymerisation" quantisation ambiguities in it if one 
works with the polynomial version and uses the natural normal ordering.
The difference between this constraint quantisation 
version of canonical quantum gravity and the LQG version is that the Fock representation 
is regular and Fock space is separable while the LQG representation is irregular and the 
LQG Hilbert space is separable. This leads to remarkable differences. 

All non Hamiltonian constraint commutators close without anomaly and 
there is also no anomaly in the commutator between non-Hamiltonian and Hamiltonian 
constraints. This is already different in the LQG representation where the 
spatial diffeomorphism constraint cannot even be defined which is the source of many 
of the complications that one faces when trying to represent the algebroid of 
quantum constraints in LQG. The commutator between Hamiltonian constraints is a priori 
ill defined because quadratic forms cannot be multiplied, however, it is conceivable 
that in the limit of vanishing regulator the commutator of regulated densely and invariantly defined 
operators converges weakly to an anomaly free quadratic form, see \cite{17} for an example.

We have also demonstrated by providing concrete examples 
that the non Hamiltonian constraints have an infinite dimensional 
kernel of generalised zero eigenvectors. On the other hand, solutions to the Hamiltonian constraints 
cannot be constructed as easily as in LQG. However, it should be noted that the known, rigorous 
solutions of the Hamiltonian constraint in LQG are in fact normalisable being in the span 
of spin network states all of whose vertices have zero volume eigenvalue or whose 
underlying graph is topologically restricted (technically, it does not contain loops of 
the kind produced by the Hamiltonian constraint). It is unclear what significance those solutions 
have in the yet to be constructed physical Hilbert space made from all solutions. In 
particular it is expected that the semiclassically relevant solutions have non-zero volume 
at each vertex and are not normalisable as the Hamiltonian constraint is an unbounded 
operator. The construction of such solutions is much more difficult in LQG than in the 
present Fock incarnation because it requires to compute matrix elements of the Hamiltonian 
constraint which is possible in closed form in Fock representations but not in LQG because 
the Hamiltonian constraint depends on the volume operator whose spectrum is not known 
in closed form (see \cite{49} for recent progress in that direction). 

We have also shown how to quantise geometrical quantities related to the quantum volume of submanifolds 
as quadratic forms in the Fock space. An important difference here with 
the LQG representation is that the form domains of these forms carry a fingerprint from 
the classical theory which rests on the silent assumption that the D-metric is nowhere 
degenerate. This can be considered as a positive feature. 
Technically this is due to the fact that in the LQG representation the D-metric 
annihilates the LQG vacuum while in the Fock representation this is not the case. This is 
also the reason why in LQG these quantities can be defined as actual operators and 
not only as quadratic forms. However, given that these quadratic forms or operators are not 
gauge invariant (they do not commute with the constraints) 
and thus do not correspond to observables, the physical significance of these 
quantities is anyway unclear.

Of more physical interest is thus the physical Hilbert space and the representation of 
quantum observables in both theories. This is very difficult to construct
in both theories because one would need to know the complete 
joint kernel of all constraints and an inner product on those. The latter cannot be 
constructed using the theory of rigged Hilbert spaces because the quantum constraints 
do not form a Lie algebra but a Lie algebroid. The probably best option in that respect is to 
pass to the master constraint \cite{25} for which rigging methods are available but 
still the computations are highly challenging in both representations. Thus 
a more promising route would therefore be the reduced phase space 
approach in which we solve all constraints classically. We have shown in this paper 
that this is straightforward 
in $D$ dimensions as far as Gauss and spatial diffeormorphism constraint are concerned 
if $D$ scalar fields are available and by using the upper triangular gauge. Even better,
this is possible using in $D=3$ just the matter content of the standard model in terms of 
six of the nine degrees of freedom as reference fields 
that reside in the massive vector boson triplet 
of the electroweak interaction of the standard model. Yet another scalar field, which 
could be the standard model Higgs field or a seventh vector boson, can then be used to solve also the Hamiltonian 
constraint in the classical theory algebraically and non-perturbatively, 
see e.g. \cite{17} for a concrete implementation using fundamental scalar fileds instead.
Thus a complete reduced phase space formulation is in principle available just using the 
matter content of the standard model. We will explore the features of the corresponding 
quantum theory, in particular its reduced Hamiltonian, in future works.

%}

\end{document}